\documentclass[prd,superscriptaddress,amsfonts,amssymb,amsmath,showpacs,twocolumn,nofootinbib]{revtex4-2}
\usepackage{bm}
\usepackage{amsfonts}
\usepackage{latexsym}
\usepackage{graphicx}
\usepackage{amsmath}
\usepackage{palatino}
\usepackage{xcolor} 
\usepackage{mathpazo}
\usepackage{adjustbox}
\usepackage{tensor}
\usepackage{rotating}
\usepackage{textcomp}
\usepackage{float}
\usepackage{booktabs}
\usepackage{dcolumn}
\usepackage[titletoc]{appendix}
\usepackage{booktabs}
\usepackage{multirow}
\usepackage{hyperref}
\hypersetup{colorlinks,citecolor=blue}
\usepackage{amsmath}
\usepackage{xcolor}
\usepackage{orcidlink}
\usepackage{epsfig}
\usepackage{caption}
\usepackage{subcaption}
\usepackage{commath}
\hypersetup{colorlinks,citecolor=blue}
\hypersetup{colorlinks=true,linkcolor=magenta,filecolor=magenta,    urlcolor=blue}

\def\be{\begin{equation}}
\def\ee{\end{equation}}
\def\bea{\begin{eqnarray}}
\def\eea{\end{eqnarray}}

\begin{document}

\title{Does DESI DR2 challenge $\Lambda$CDM paradigm ?}

\author{Himanshu Chaudhary}
\email{himanshu.chaudhary@ubbcluj.ro,\\
himanshuch1729@gmail.com}
\affiliation{Department of Physics, Babeș-Bolyai University, Kogălniceanu Street, Cluj-Napoca, 400084, Romania}
\author{Salvatore Capozziello}
\email{capozziello@na.infn.it}
\affiliation{Dipartimento di Fisica ``E. Pancini", Universit\`a di Napoli ``Federico II", Complesso Universitario di Monte Sant’ Angelo, Edificio G, Via Cinthia, I-80126, Napoli, Italy,}
\affiliation{Istituto Nazionale di Fisica Nucleare (INFN), sez. di Napoli, Via Cinthia 9, I-80126 Napoli, Italy,}
\affiliation{Scuola Superiore Meridionale, Largo S. Marcellino, I-80138, Napoli, Italy.}
\author{Vipin Kumar Sharma}
\email{vipinkumar.sharma@iiap.res.in}
\affiliation{Indian Institute of Astrophysics, Koramangala II Block, Bangalore 560034, India}
\author{Ghulam Mustafa}
\email{gmustafa3828@gmail.com}
\affiliation{Department of Physics,
Zhejiang Normal University, Jinhua 321004, People’s Republic of China}

\begin{abstract}
Although the debate about the systematic errors of DESI DR1 is still open, recent DESI DR2 is consistent with DESI DR1 and further strengthens the results of DESI DR1. In our analysis, both the LRG1 point at $z_{\mathrm{eff}}=0.510$ and the LRG3+ELG1 point at $z_{\mathrm{eff}}=0.934$ are in tension with the $\Lambda$CDM-anchored value of $\Omega_m$ inferred from Planck and the SNe Ia compilations Pantheon$^{+}$, Union3, and DES-SN5YR. For LRG1 the tensions are $2.42\sigma$, $1.91\sigma$, $2.19\sigma$, and $2.99\sigma$, respectively; for LRG3+ELG1 they are $2.60\sigma$, $2.24\sigma$, $2.51\sigma$, and $2.96\sigma$, respectively. From low to high redshift bins, DESI DR2 shows improved consistency relative to DESI DR1: the $\Omega_m$ tension decreases from $2.20\sigma$ to $1.84\sigma$. However, DESI DR2 alone does not provide decisive evidence against the $\Lambda$CDM model, and the apparent signal is largely driven by specific tracers, LRG1 and LRG2. In the $\omega_0\omega_a$CDM analysis, including all tracers yields a posterior mean with $\omega_0>-1$, which aligns with scenarios of dynamical dark energy as a potential explanation and suggests that the DESI DR2 challenges $\Lambda$CDM paradigm. While removing LRG1 and/or LRG2 fully restores $\Lambda$CDM concordance (i.e., $\omega_0\to -1$); we also find $\omega_0^{\text{(LRG1)}} > \omega_0^{\text{(LRG2)}}$, indicating LRG1 drives the apparent dynamical dark energy trend more strongly. Model selection using the natural log Bayes factor $\ln\mathrm{BF}\equiv \ln(\mathcal{Z}_{\Lambda\mathrm{CDM}}/\mathcal{Z}_{\omega_0\omega_a\mathrm{CDM}})$ shows weak evidence for $\Lambda$CDM when LRG1, LRG2, or both are removed, and it is inconclusive for the full sample; thus, the data do not require the extra $\omega_a$ freedom, and the apparent $\omega_0>-1$ preference should be interpreted cautiously as a manifestation of the $\omega_0$$\omega_a$ degeneracy under limited per tracer information.
\end{abstract}

\maketitle


\section{Introduction}\label{sec_1}
Many different observational probes and experiments have been conducted over the years in order to  interpret  the dark energy fluid \citep{SupernovaSearchTeam:1998fmf,SupernovaCosmologyProject:1998vns,Bamba:2012cp}. A comprehensive analysis of cosmological data sets at different redshifts is paramount to understanding its properties, with a rule of thumb of testing as many ranges as possible. Whenever both BAO (Baryon Accoustic Oscillations) and SNe deviate significantly from $\Lambda$CDM predictions at the same redshift range, this provides strong phenomenological evidence that dark energy is dynamic rather than a cosmological constant \citep{Sousa-Neto:2025gpj,Notari:2024zmi,Sharma:2025qmv}.

The $\Lambda$CDM model is widely regarded as the best model for explaining most cosmological observations despite its degeneracies at  theoretical level \citep{Weinberg:1988cp,YaBZeldovich_1968}. However, using their first year data, the DESI collaboration \citep{DESI:2024mwx,DESI:2024aqx,karim2025desi,DESI:2024kob} has found that dark energy is evolving at a $\geq3\sigma$ significance level.  This result will inevitably bring the shifts of the bestfit values of relevant $\Lambda$CDM cosmological parameters. 
Therefore, modeling the dark energy through parameterization is one of the most direct ways to understand its dynamical behavior in a model-independent approach \citep{Sahni:2006pa,Chevallier:2000qy,Linder:2002et,Park:2024pew,wolf2025matching,wolf2025robustness}. This makes model-independent techniques an attractive route to discover dark energy properties at a deeper level \citep{Vilardi:2024cwq}. In this paper, we want to perform our analysis for different redshifts range and compare the DESI DR1 and DR2 predictions for $\Lambda$CDM model and $\omega_0\omega_a$CDM model.

Our paper is organised as follows. In Section \ref{sec_2}, we introduce the cosmological background equations and models.
Section \ref{sec_3} details the core of this work with datasets and methodology using Markov Chain Monte Carlo (MCMC) sampling against the publicly available DESI DR2 data, while Section \ref{sec_4} is dedicated to the discussion of results.  We discuss the implications of our results and conclusions in Section \ref{sec_4}.

\section{Cosmological Background and Models}\label{sec_2}

Considering the spatially flat Friedmann-Lema\^{i}tre-Robertson-Walker (FLRW) universe at relatively late times, that is  $z<< 10^2$, where the density of radiation can be safely ignored; 
the first Friedmann equation  reads:
\begin{equation}
    H^2 = \frac{8\pi G}{3} \left(\rho_\textrm{m,0} a^{-3} +  \rho_{\textrm{de}} \right).
    \label{eqn:Hubble_1}
\end{equation}
The continuity equation in FLRW is,
\begin{equation}\label{eq_2}
\dot{\rho}_{\text{x}} + 3H(1 + w_{\text{x}}) \rho_{\text{x}} = 0,
\end{equation}
where \(\rho_\text{x}\) represents the energy density of each component with x$\in$ (de, m), over $(\dot{})$ represents the cosmic time derivative, and $w_x$ represents the equation of state parameter (EoS). Here "de" and "m" are the dark energy and the matter components respectively.

Suppose that we parameterize $w(a)$, so that the evolution of dark energy ($\rho_{\text\rm{de}}$) within \eqref{eqn:Hubble_1} is the following solution of \eqref{eq_2}
\begin{equation}\label{eqn:rhode}
\rho_{\textrm{de}} = \frac{\rho_{\textrm{de,0}}}{a^3} \exp\left(-3\int^a_1 \frac{w_{\textrm{de}}(a')}{a'} \, da'\right),
\end{equation}
where $\rho_{\rm de, 0}$ is the present value of the dark energy density. With  a particular form choice $w_\text{de}(a)$, \eqref{eqn:Hubble_1}, it can be used to determine the cosmological evolution. 

In principle, there is no dictum that specifies the best parameterisations. Nevertheless, by using observational data, it is possible to find parameterisations that are cosmologically viable.  Let us now outline the $\Lambda$CDM model and its straightforward dynamic improvement represented by the  $w_0 w_a$CDM model. \\  \\

$1.$ \textbf{\( \Lambda \)CDM Model }\\

 This is the Standard Concordance Cosmological Model characterized by constant EoS, \( w_{\text{de}} = -1 \).   Eq.\eqref{eqn:rhode}  gives $\rho_\text{de}=\rho_\text{de,0}$.  
The dimensionless Hubble function for flat FLRW universe reads:
\begin{equation}
E(z)^2  \left( \equiv \frac{H(z)}{H_0} \right)^2 = \Omega_m(1 + z)^3 + (1 - \Omega_m). \label{eqn:Hubble_2}
\end{equation}


2. \textbf{$\omega_0 \omega_a$CDM Model}\\


It is possible to approximate the behavior of several dark energy models by using the function $w_{\rm de}$ in a model-independent way but  time-dependent parameterizations.

Specifically, we consider the two-parameter model ($\omega_0\omega_a$CDM) with the following free parameters: $\omega_0$, which represents the current value of $w_{\rm de}(a)$, and $w_a = -\frac{dw_{\rm de}(a)}{da}\big|_{a = a_0}$, which quantifies dynamical characteristics of $\omega_{\rm de}(a)$. The standard $\Lambda$CDM model can be recovered for the choices $\omega_0 = -1$ and $\omega_a = 0$. In this aspect, model-independent techniques applied to observational data provide an interesting approach to understanding dark energy  properties in more detail. To this category belong the popular CPL ansatz \citep{Chevallier:2000qy,Linder:2002et,Chudaykin:2020ghx,Park:2024pew} for the dark energy equation of state (EoS), defined as

\begin{equation}
\omega(z) = \omega_0 + \frac{z}{1 + z} \omega_a=\omega_0+\omega_a(1-a). 
\label{w}
\end{equation}
Eq. \eqref{eqn:rhode} gives,
\begin{equation}
\rho_\text{de}=\rho_\text{de,0}(1+z)^{3(1+w_0+w_a)}\exp{(-\frac{3w_az}{1+z})}
\label{rho(z)}
\end{equation}
The corresponding Hubble function is given by \citep{Escamilla-Rivera:2019aol}:

\begin{equation}
E(z)^2 = \Omega_m(1 + z)^3 + \Omega_x(1 + z)^{3(1 + \omega_0 + \omega_a)} e^{-\frac{3\omega_a z}{1 + z}}. 
\label{E(z)}
\end{equation}

The $\omega_0 \omega_a$CDM model is widely used because of its flexibility and robust behaviour in describing the evolution of dark energy.

The next section scrutinizes both models based on the DESI DR2 data sets.

\section{Dataset and Methodology}\label{sec_3}
Let us consider now the recent measurements of baryon acoustic oscillations (BAO) from more than 14 million galaxies and quasars taken from the Dark Energy Spectroscopic Instrument (DESI) Data Release 2 (DR2) \footnote{\url{https://github.com/CobayaSampler/bao_data}} \cite{karim2025desi}. These data have been extracted from various tracers, including BGS, Luminous Red Galaxies (LRG1–3), Emission Line Galaxies (ELG1, ELG2), Quasi-Stellar Objects (QSOs), and Lyman-$\alpha$ forests.

To analyze these data, we employ the nested sampling algorithm, a Bayesian inference technique well-suited for navigating complex, high-dimensional parameter spaces. Nested sampling transforms the challenging multi-dimensional integration of the Bayesian evidence into a more manageable one dimensional integral over the prior volume. This method not only enables efficient parameter estimation but also facilitates the computation of  Bayesian evidence, $\mathcal{Z} = p(D|M)$, which is crucial for model comparison.

For the nested sampling implementation, we use the \textsc{PyPolyChord} library\footnote{\url{https://github.com/PolyChord/PolyChordLite}} \cite{handley2015polychord1,handley2015polychord2}, which is optimized for high-dimensional spaces and can handle multimodal posterior distributions through clustering techniques. In our analysis, we apply uniform priors to the model parameters, which are implemented via PyPolyChord's \texttt{UniformPrior} class. We consider two cosmological models: the standard $\Lambda$CDM model and the $\omega_0\omega_a$CDM model. For both models, we specify appropriate priors on their parameters. In the case of the $\Lambda$CDM model, we use uniform priors on the following parameters: $H_0 \in [50, 100]$ km s$^{-1}$ Mpc$^{-1}$, $\Omega_{m0} \in [0, 1]$, and $r_d \in [100, 200]$ Mpc. For the $\omega_0 \omega_a$CDM model, we use the following uniform priors $\omega_0 \in [-3, 1]$ and $\omega_a \in [-3, 2]$, with the additional condition that $\omega_0 + \omega_a < 0$.

To efficiently explore the parameter space and detect multiple posterior modes, we configure PyPolyChord to use 300 live points with clustering enabled. For analyzing and visualizing the results, we utilize the \texttt{getdist} package\footnote{\url{https://github.com/cmbant/getdist}} \cite{lewis2019getdist}, which generates detailed marginalized posterior distributions and parameter correlation plots.
To analyze the BAO dataset, we need to compute two important distance measures: the Hubble distance 
\begin{equation}
D_H(z) = \frac{c}{H(z)}\,,
\end{equation}
and the comoving angular diameter distance 
\begin{equation} D_M(z) = \frac{c}{H_0} \int_0^z \frac{dz'}{H(z')}\,,
\end{equation}
, and the volume-averaged distance 
\begin{equation}
D_V(z) = \left[ z D_M^2(z) D_H(z) \right]^{1/3}
\end{equation}.
These distances are then used to calculate the ratios $D_H(z)/r_d$, $D_M(z)/r_d$, and $D_V(z)/r_d$  which we compare directly with observational data. These calculated values are then compared with the predictions made by the observations.

Furthermore, we also calculate the following ratio: 
\begin{equation}
\frac{D_M / r_d}{D_H / r_d} = \int_0^z \frac{1}{E(z')} \, dz'.
\end{equation} 
This approach is independent of the sound horizon $r_d$. The ratio $\frac{D_M / r_d}{D_H / r_d}$ involves the angular diameter distance $D_M$ and the Hubble distance $D_H$, both normalized by $r_d$. This ratio depends on the function $E(z)$. The dataset for $\frac{D_M}{D_H}$ can be found in Table 4, Column 6 of \cite{karim2025desi}, corresponding to each tracer. For each tracer, one can compute the model value $\frac{D_M}{D_H}_{\text{Model}}$ and compare it with the observational dataset using the standard likelihood function. The likelihood function to compare the model prediction $\frac{D_M}{D_H}_{\text{Model}}$ with the observational data $\frac{D_M}{D_H}_{\text{Obs}}$ is given by: 
\begin{equation}
\mathcal{L}(\theta) = \prod_{i} \exp\left[-\frac{1}{2} \left(\frac{\frac{D_M}{D_H}_{\text{Obs},i} - \frac{D_M}{D_H}_{\text{Model},i}(\theta)}{\sigma_i}\right)^2\right],
\end{equation} 
where: $\frac{D_M}{D_H}_{\text{Obs},i}$ is the observed ratio for the $i$-th tracer, $\left(\tfrac{D_M}{D_H}\right)_{\text{Model},i}(\boldsymbol{\theta})$ is the predicted ratio for the $i$ th tracer based on the cosmological model with parameter set $\boldsymbol{\theta} = \{\Omega_m, H_0, \omega_0, \omega_a, r_d, \ldots\}$, and $\sigma_i$ is the uncertainty associated with the $i$-th observed data point. It is important to note that, in our analysis, $r_d$ is treated as a free parameter \cite{pogosian2020recombination,jedamzik2021reducing,pogosian2024consistency,lin2021early,vagnozzi2023seven}. We then construct the derived quantity $h\,r_d$ (with $h \equiv H_0/100$) and obtain its posterior by propagating the MCMC samples of $h$ and $r_d$ sample by sample.

\section{Results}\label{sec_4}
\begin{figure}
\centering
\includegraphics[scale=0.44]{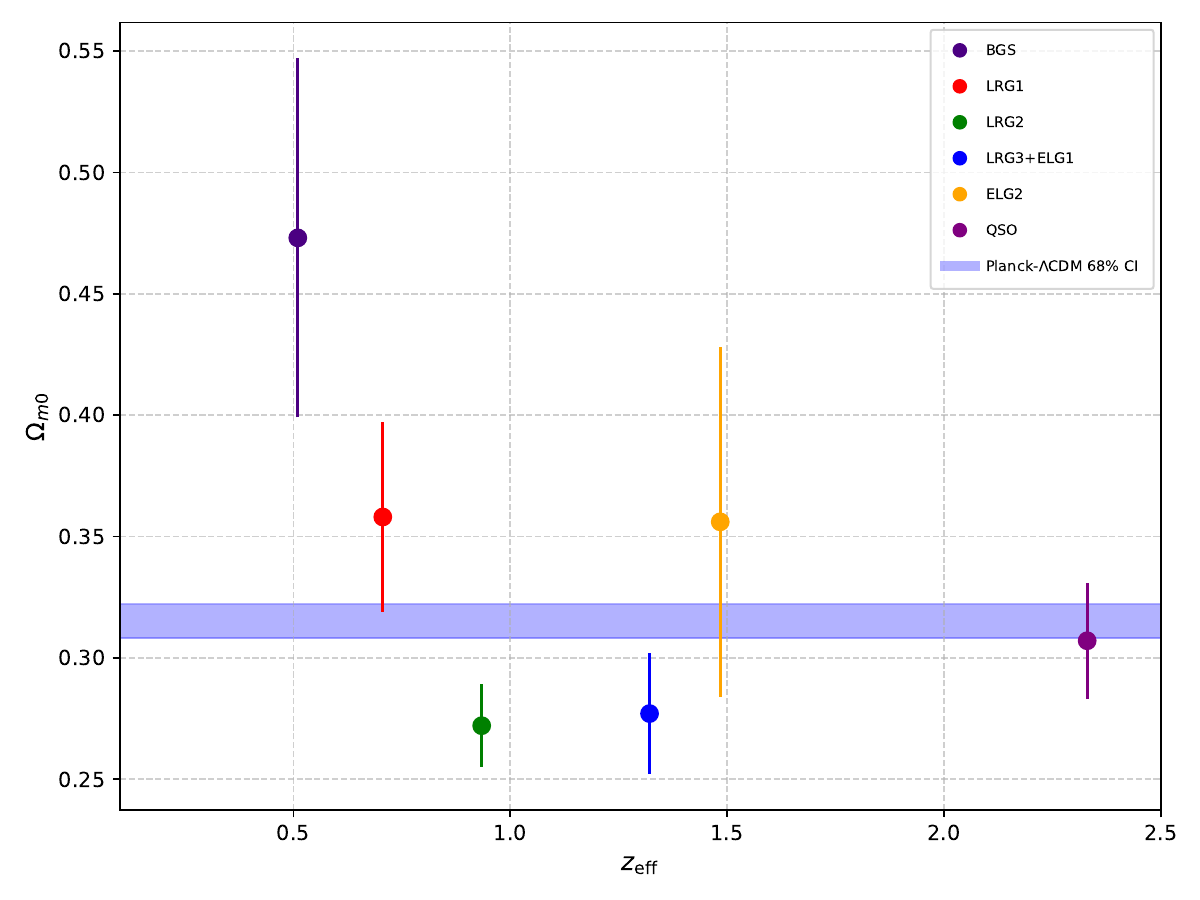}
\caption{The figure shows the values of $\Omega_m$ at 68\% confidence intervals for different $z_{\text{eff}}$ values. The blue band represents the Planck $\Lambda$CDM prediction for $\Omega_m$.}\label{fig_1}
\end{figure}

\begin{figure}
\centering
\includegraphics[scale=0.6]{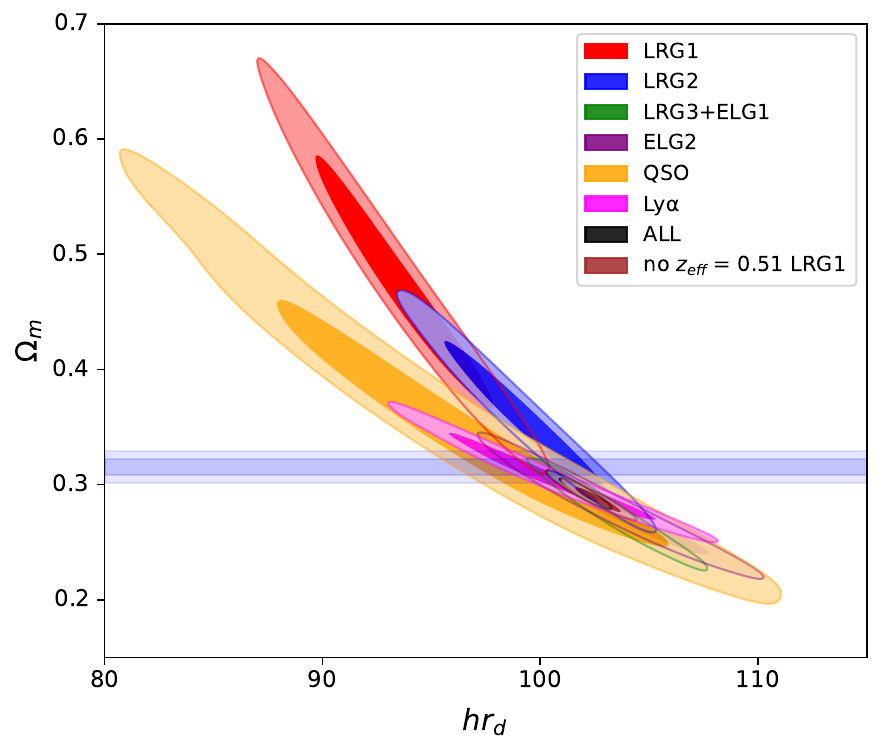}
\caption{The figure shows the posterior distributions of different tracers corresponding to different $z_{\text{eff}}$ from the DESI DR2 dataset within the $\Lambda$CDM model. These are presented at 68\% (1$\sigma$) and 95\% (2$\sigma$) confidence levels in the $\Omega_m - h r_d$ contour plane. The blue band represents the Planck $\Lambda$CDM prediction for $\Omega_m$.}\label{fig_2}
\end{figure}

Fig.~\ref{fig_1} shows the values of $\Omega_m$ at different $z_{\text{eff}}$ values. We observe that the predicted value of $\Omega_m$ using the LRG1 dataset, corresponding to $z_{\text{eff}} = 0.51$, deviates from the Planck-$\Lambda$CDM confidence interval represented by the blue band. Fig.~\ref{fig_2} shows the posterior distributions of different tracers at 1$\sigma$ and 2$\sigma$ confidence intervals, obtained from various $z_{\text{eff}}$ measurements using DESI DR2 datasets. Notably, one can observe that the contours for the LRG1 dataset do not lie within the blue band, which represents the Planck $\Lambda$CDM prediction for $\Omega_m$. 

Table~\ref{tab_1} highlights two main points. First, the LRG1 data at $z_{\rm eff} = 0.510$ yields an unexpectedly high value of $\Omega_m$ compared to the Planck prediction, $\Omega_m = 0.315 \pm 0.007$ \cite{aghanim2020planck}. In DESI DR2, LRG1 predicts $\Omega_m = 0.473 \pm 0.065$, corresponding to a $\sim 2.42\sigma$ discrepancy with the Planck prediction. Other redshift bins also show tensions with the Planck prediction: LRG2 ($z_{\rm eff} = 0.760$) predicts $\Omega_m = 0.358 \pm 0.032$ ($1.31\sigma$), LRG3+ELG1 ($z_{\rm eff} = 0.934$) gives $\Omega_m = 0.272 \pm 0.015$ ($2.60\sigma$), ELG2 ($z_{\rm eff} = 1.321$) yields $\Omega_m = 0.277 \pm 0.021$ ($1.72\sigma$), QSO ($z_{\rm eff} = 1.484$) gives $\Omega_m = 0.356 \pm 0.061$ ($0.67\sigma$), and Ly$\alpha$ ($z_{\rm eff} = 2.330$) shows only a $0.36\sigma$ tension. These findings indicate that the LRG1 discrepancy at $z_{\rm eff} = 0.510$ is not unique, as other tracers most notably LRG3+ELG1 at $z_{\rm eff} = 0.934$ show even stronger deviations from the Planck prediction. More broadly, all tracers except Ly$\alpha$ exhibit some level of tension, with a pattern that is complex and not monotonic in redshift. A similar trend is observed in Table~1 of \cite{colgain2024does}, where the predicted values of $\Omega_m$ in different redshift bins also exhibit discrepancies with the Planck prediction.

Consequently, these values also show discrepancies with various SNe Ia measurements: Pantheon$^+$ ($\Omega_m = 0.334 \pm 0.018$) \cite{brout2022pantheon}, Union3 ($\Omega_m = 0.356^{+0.028}_{-0.026}$) \cite{rubin2025union}, and DES-SN5YR ($\Omega_m = 0.352 \pm 0.017$) \cite{abbott2024dark}. For DESI DR2, the LRG1 sample exhibits tensions of approximately $2.06\sigma$, $1.67\sigma$, and $1.80\sigma$ with these three datasets, respectively. In comparison, the LRG3+ELG1 sample shows even larger tensions of about $2.24\sigma$, $2.51\sigma$, and $2.96\sigma$. Note that these SNe Ia samples typically have low effective redshifts around $z_{\text{eff}} \sim 0.3$.

This trend extends beyond the Planck and SNe Ia calibrations. At $z_{\rm eff} = 0.510$, SDSS-IV data predict $\Omega_m \sim 0.340 \pm 0.09$ \cite{o2022revealing}, while DESI DR1 predicts $\Omega_m \sim 0.67_{-0.17}^{+0.18}$ \cite{colgain2024does}. In comparison, the DESI DR2 measurement at the same redshift shows a tension of approximately $1.20\sigma$ with SDSS-IV and $1.08\sigma$ with DESI DR1, indicating a deviation across different surveys. A similar discrepancy is observed within the DESI DR1 and DESI DR2 measurements at $z_{\rm eff} = 0.706$. Using DESI DR1 data, \cite{o2022revealing} predicted $\Omega_m = 0.219_{-0.069}^{+0.087}$ (see Table 1). However, DESI DR2 measurements at the same redshift give $\Omega_m \sim 0.358 \pm 0.032$, showing an increase of approximately $1.83\sigma$. These predictions also show a similar discrepancy when compared to the value $\Omega_m = 0.49 \pm 0.11$ predicted by \cite{colgain2024does} at $z_{\rm eff} = 0.706$ using SDSS-IV. The comparison between DESI DR1 and this value shows a discrepancy of $2.09\sigma$, while the comparison between DESI DR2 and \cite{colgain2024does} shows a discrepancy of $1.14\sigma$. It is worth noting that DESI DR1 and DESI DR2 agree with the predicted value of $\Omega_m$ at $z_{eff} = 0.934$ and only show a tension of about $0.10\sigma$.

The second notable feature is the negative correlation between $\Omega_m$ and $h r_d$, as shown in Fig.~\ref{fig_2}, which shows the $\Omega_m$–$h r_d$ contour planes for the various DESI DR2 tracers. This negative correlation has already been well documented in different BAO analyses \cite{eisenstein2005detection,percival2010baryon,aubourg2015cosmological,alam2017clustering,alam2021completed}, observational Hubble data (OHD) \cite{colgain2024putting,dainotti2022evolution}, Type Ia Supernovae (SNe Ia) \cite{o2022revealing,dainotti2021hubble,colgain2024putting,jia2023evidence,pasten2023testing,malekjani2024redshift,wagner2022solving,hu2022revealing,dainotti2023hubble}, combinations of OHD and SNe Ia \cite{colgain2024putting,krishnan2020there}, Gamma-Ray Bursts (GRBs) \cite{dainotti2022gamma, Bargiacchi:2024srw}, standardizable QSOs \cite{o2022revealing,colgain2024putting,risaliti2019cosmological,lusso2020quasars,dainotti2023quasars,dainotti2022quasar,bargiacchi2023gamma,pourojaghi2022can}, and discussions on strong lensing time delays in lensed QSOs \cite{wong2020h0licow,shajib2020strides,millon2020tdcosmo} and SNe Ia \cite{kelly2023constraints,pascale2025sn}. This is clearly shown in Fig. 5 of \cite{millon2020tdcosmo} (see also \cite{li2024determining}), where the error bars for $H_0$ from SN Refsdal and SN H0pe do not overlap, indicating a disagreement of about $1.5\sigma$. It's important to note that SN H0pe has a lens redshift of $z = 0.35$, while SN Refsdal has a lens redshift of $z = 0.54$. This makes the trend observed in Figure 5 of \cite{millon2020tdcosmo} consistent with the decreasing trend of $H_0$ with lens redshift, as originally reported the appendix of \cite{wong2020h0licow}.

A growing body of work is either questioning \cite{khadka2020using,khadka2021determining,singal2022x,petrosian2022can,zajavcek2024effect} or improving the Risaliti Lusso standardizable QSO prescription \cite{dainotti2022quasar,dainotti2024scavenger,dainotti2024new, Bargiacchi:2021hdp, Benetti:2025ljc}. Despite these corrections, residual evolution in the $\Omega_m$ parameter is still reported \cite{dainotti2024scavenger,dainotti2024new,lenart2023bias,camarena2025designing}. Additionally, DES SNe, which have a higher effective redshift \cite{abbott2024dark}, show a larger $\Omega_m$, consistent with these observations. Moving beyond traditional probes like SNe, QSOs \cite{risaliti2019cosmological,lusso2020quasars} and GRBs \cite{demianski2017cosmology,demianski2017cosmology,dainotti2023gamma,khadka2021gamma,alfano2024cosmological,dainotti2008time,srinivasaragavan2020investigation} calibrated by SNe may also yield higher $\Omega_m$ values at higher redshifts. This could be due to large data scatter rather than any variation in $\Lambda$CDM parameters. Comparing SNe in overlapping redshift ranges can help distinguish between these possibilities \cite{o2022revealing}. Returning to the main point, each $z_{\rm eff}$ from DESI DR2 further supports the negative correlation between $\Omega_m$ and $h r_d$, which is also evident in Fig.~\ref{fig_2}, where the posterior distributions for each tracer are tilted from the top left to the bottom right.
\begin{table}
\centering
\begin{tabular}{c|c|c|c}
\hline
\textbf{Tracer} & \textbf{$z_{\text{eff}}$} & \textbf{$h r_{d}$} & \textbf{$\Omega_{m}$} \\
\hline
LRG1 & 0.510 & $93.64_{\pm 2.703}^{\pm 5.502}$ & $0.473_{\pm 0.065}^{\pm 0.132}$ \\
LRG2 & 0.706 & $99.24_{\pm 2.207}^{\pm 4.420}$ & $0.358_{\pm 0.032}^{\pm 0.075}$ \\
LRG3+ELG1 & 0.934 & $103.50_{\pm 1.558}^{\pm 3.162}$ & $0.272_{\pm 0.015}^{\pm 0.036}$ \\
ELG2 & 1.321 & $103.56_{\pm 2.637}^{\pm 5.262}$ & $0.277_{\pm 0.021}^{\pm 0.047}$ \\
QSO & 1.484 & $96.04_{\pm 6.197}^{\pm 12.206}$ & $0.356_{\pm 0.061}^{\pm 0.130}$ \\
Ly$\alpha$ & 2.330 & $100.29_{\pm 3.201}^{\pm 5.871}$ & $0.307_{\pm 0.021}^{\pm 0.046}$ \\
ALL & - & $101.79_{\pm 0.611}^{\pm 1.172}$ & $0.297_{\pm 0.007}^{\pm 0.013}$ \\
No $z_{\text{eff}} = 0.510$ LRG1 & - & $102.59_{\pm 0.700}^{\pm 1.449}$ & $0.287_{\pm 0.007}^{\pm 0.013}$ \\
\hline
\end{tabular}
\caption{The table shows the values of $\Omega_m$ and $h r_d$ at 68\% (1$\sigma$) and 95\% (2$\sigma$) confidence intervals, obtained from different tracers using DESI DR2 measurements. (Note: In this table, we do not consider the BGS data point.)}
\label{tab_1}
\end{table}

\begin{figure}
\centering
\includegraphics[scale=0.6]{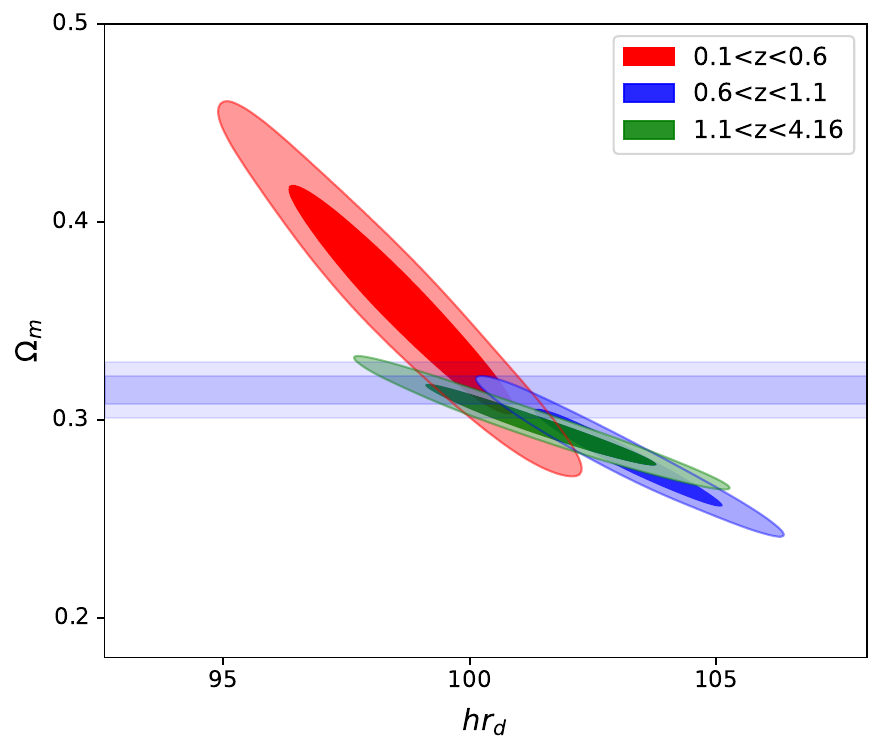}
\caption{The figure shows the posterior distributions at 68\% (1$\sigma$) and 95\% (2$\sigma$) in the $\Omega_m - h r_d$ plane using different redshift bins from the DESI DR2 compilation. The blue band represents the Planck $\Lambda$CDM prediction for $\Omega_m$.}\label{fig_3}
\end{figure}

In Fig~\ref{fig_3}, we show the posterior distributions of the $\Omega_m - hr_d$ plane at the 1$\sigma$ and 2$\sigma$ confidence levels, using different redshift bins from the DESI DR2 compilation. Table~\ref{tab_2} shows the corresponding numerical values obtained for $\Omega_m$ and $h r_d$ at those redshift bins. One can observe that in the $\Omega_m - h r_d$ plane, the $\Lambda$CDM yield different preferred values of $\Omega_m$ across redshift bins, without a clear monotonic pattern. These variations are more naturally interpreted as statistical fluctuations and parameter degeneracies than as evidence for a true physical redshift evolution of $\Omega_m$. The corresponding numerical values can be seen in the third column of Table~\ref{tab_2}. Statistically speaking, the matter density parameter $\Omega_m$ shows a discrepancy of approximately 1.84 $\sigma$ when moving from the low redshift bin to the higher redshift bin, and a discrepancy of 0.78 $\sigma$ when moving from the high redshift bin to an even higher redshift bin. On the other hand, in the same plane, the product $h r_d$ initially predicts low values at low redshift bins, increases to intermediate values in intermediate redshift bins, and then decreases again at larger redshift bins. Indeed, $\Omega_m$ and $h r_d$ exhibit a negative correlation.

The DESI DR2 dataset shows some improvement when moving from higher to even higher redshifts, compared to the DESI DR2 predictions in \cite{colgain2024does} (see Table 2). The authors computed the value of $\Omega_m$ at different redshift bins and found that at low redshift, they predict $\Omega_m = 0.665 \pm 0.188$. At higher redshifts, they predict $\Omega_m = 0.231 \pm 0.033$, and at even higher redshift bins, they found $\Omega_m = 0.324 \pm 0.044$. Statistically, the matter density at low redshift shows a tension of about 2.20 $\sigma$, and when moving from the high to even higher redshift bins, the tension is about 1.85 $\sigma$.

Note that at around 2$\sigma$, one might consider disregarding these results. However, it is well established that the matter density in the $\Lambda$CDM Universe is approximately 30\%. Based on the DESI DR2 results, there is now a region of the Universe, particularly at lower redshifts, where the matter density reaches 47.3\%, compared to about 65\% in light of the DESI DR1 predictions. This is strikingly inconsistent with the standard model.
\begin{table}
\centering
\begin{tabular}{c|c|c|}
\hline
\textbf{Redshift Range (z)} & \textbf{$H_{0} r_{d}[100 \, \mathrm{km}/\mathrm{s}]$} & \textbf{$\Omega_{m}$} \\
\hline
$0.1 < z < 0.6$ & $93.50_{\pm 1.588}^{\pm 3.021}$ & $0.362_{\pm 0.041}^{\pm 0.073}$ \\
$0.6 < z < 1.1$ & $103.16_{\pm 1.254}^{\pm 2.465}$ & $0.281_{\pm 0.016}^{\pm 0.030}$ \\
$1.1 < z < 4.16$ & $101.45_{\pm 1.454}^{\pm 3.033}$ & $0.297_{\pm 0.013}^{\pm 0.024}$ \\
\hline
\end{tabular}
\caption{This table shows the values of $\Omega_m$ at 68\% (1$\sigma$) and 95\% (2$\sigma$) confidence for different redshift bins, derived from DESI DR2 measurements.}
\label{tab_2}
\end{table}

In our analysis, using the full DESI DR2 compilation, we obtain $\Omega_m = 0.297 \pm 0.007$. This value is in tension with Planck, Pantheon$^+$, Union3, and DES-SN5YR at approximately $1.82\sigma$, $1.92\sigma$, $2.19\sigma$, and $2.99\sigma$, respectively. While all of these differences remain below the $3\sigma$ threshold, they are nonetheless significant and warrant further investigation. When we exclude the LRG1 tracers, we obtain a lower value of $\Omega_m = 0.289 \pm 0.007$, which increases the tensions to $2.83\sigma$, $2.43\sigma$, $2.56\sigma$, and $3.54\sigma$ with Planck, Pantheon$^+$, Union3, and DES-SN5YR, respectively. These results clearly show that removing LRG1 increases the tension with Planck from $1.82\sigma$ to $2.83\sigma$, indicating that LRG1 actually pulls the combined result closer to the Planck prediction rather than driving it away.

The assumption that $\Omega_m$ is constant in $\Lambda$CDM is challenged by mild tensions across tracers, with some approaching the $3\sigma$ level. Given the underconstrained nature of tracer wise fits, such tensions should be interpreted with caution, as they may arise from systematic effects or model degeneracies rather than genuine deviations from $\Lambda$CDM. This issue must be understood within a broader context, alongside the existing tension in $H_0$, which suggests a potential problem in the background Cosmology. There is also the tension in $S_8 = \sigma_8 \sqrt{\Omega_m / 0.3}$. While these may seem like separate issues, it is crucial to recognize that they could be interconnected, especially if discrepancies in $\Omega_m$ are confirmed \cite{akarsu2024lambda}. The reason is that $H_0$ is correlated with $\Omega_m$ at the background level in the late Universe, and $S_8$ is clearly dependent on $\Omega_m$. Therefore, if $\Omega_m$ is not constant in the $\Lambda$CDM model, the tensions in $H_0$ and $S_8$ are likely symptoms of the same underlying issue.
\begin{table}
\centering
\begin{tabular}{c|c|c}
\hline
\textbf{Tracer} & \textbf{$z_{\text{eff}}$} & \textbf{$\Omega_{m}$} \\
\hline
LRG1 & 0.510 & $0.485{\pm 0.011}$ \\
LRG2 & 0.706 & $0.363{\pm 0.054}$ \\
LRG3+ELG1 & 0.934 & $0.273{\pm 0.026}$ \\
ELG2 & 1.321 & $0.280{\pm 0.032}$ \\
QSO & 1.484 & $0.389{\pm 0.014}$ \\
Ly$\alpha$ & 2.330 & $0.309{\pm 0.031}$ \\
ALL & - & $0.299{\pm 0.013}$ \\
No $z_{eff}= 0.51$ LRG1 & - & $0.292{\pm 0.012}$ \\
\hline
\end{tabular}
\caption{Values of $\Omega_m$ for different tracers at various effective redshifts $z_{\text{eff}}$, independent of the $hr_d$ dependence.}
\label{tab_3}
\end{table}
\begin{figure}
\centering
\includegraphics[scale=0.38]{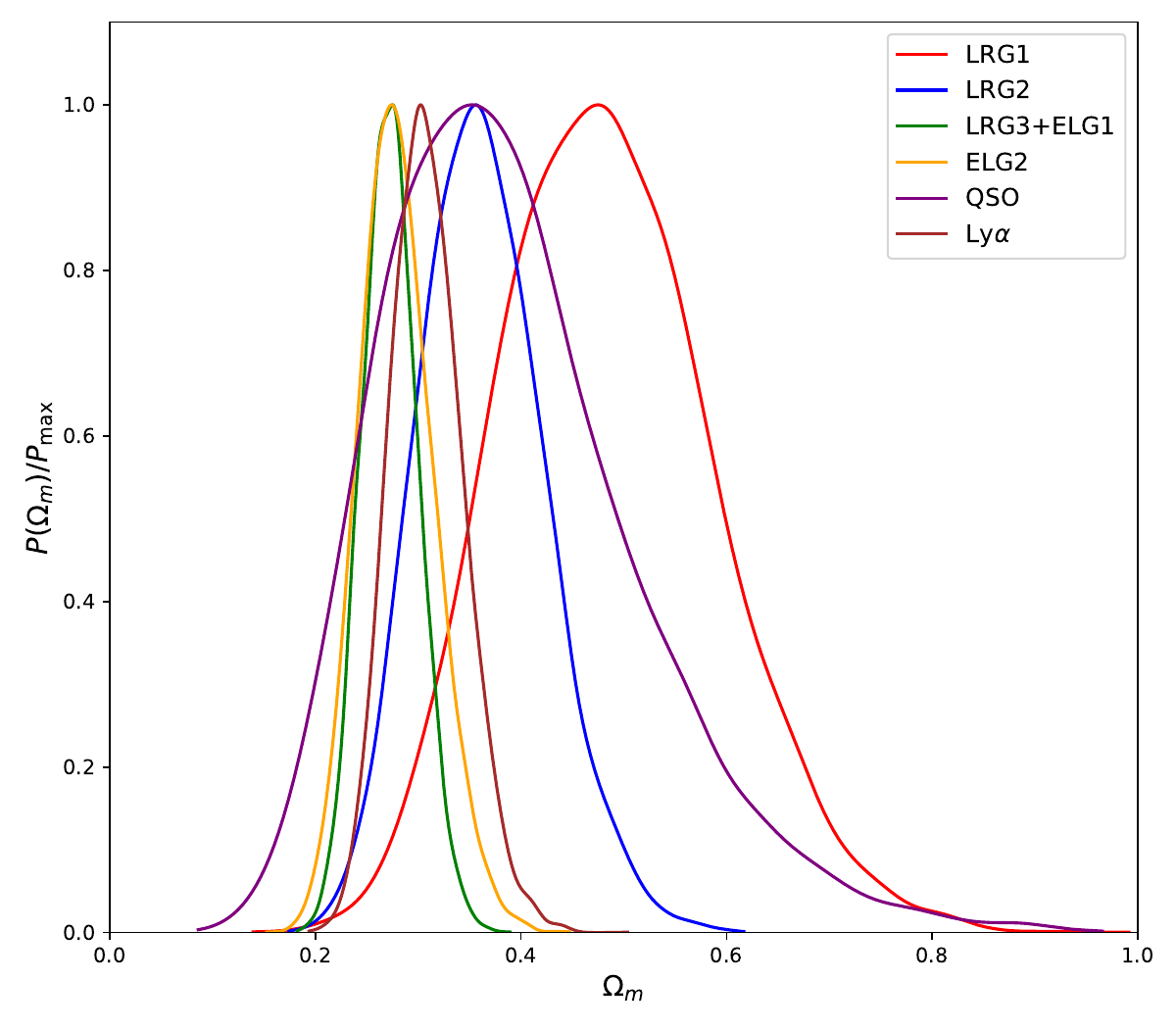}
\caption{The figure shows the normalized probability distribution of $\Omega_m$ for different tracers at various effective redshifts $z_{\text{eff}}$, independent of the $h r_d$ dependence}\label{fig_4}
\end{figure}
Table~\ref{tab_3} shows the values of $\Omega_m$ using the $D_M/D_H$ measurements at different BAO tracers. This approach is intended to remove the dependence on $H_0r_d$, allowing for a more direct constraint on $\Omega_m$. It was expected that this method would yield smaller errors; however, this was not the case. We again observe an anomaly in the DESI DR2 LRG1 data, as evident in Fig~\ref{fig_4}. Specifically, the LRG1 data at $z_{\text{eff}}$ yields a higher value of $\Omega_m$, indicating that the issue continues.

From the above discussion, it has been confirmed that the value of $\Omega_m$ at $z_{\text{eff}} = 0.51$ using the LRG1 dataset shows a strong disagreement with the Planck-$\Lambda$CDM model, with a discrepancy of about 2.12$\sigma$. It has also been shown that this constraint on $\Omega_m$ disagrees with the Pantheon$^+$ compilation at the 2.3$\sigma$ level, despite Type Ia Supernovae being highly sensitive to similar redshift ranges. These disagreement between the DESI and Pantheon$^+$ SNe Ia datasets along with other SNe Ia datasets, which all generally agree on $\Omega_m \sim 0.3$ at lower redshifts calls for further investigation.
\subsection{Constraints on \( \omega_0 \omega_a \)CDM Model}
\begin{figure}
\centering
\includegraphics[scale=0.6]{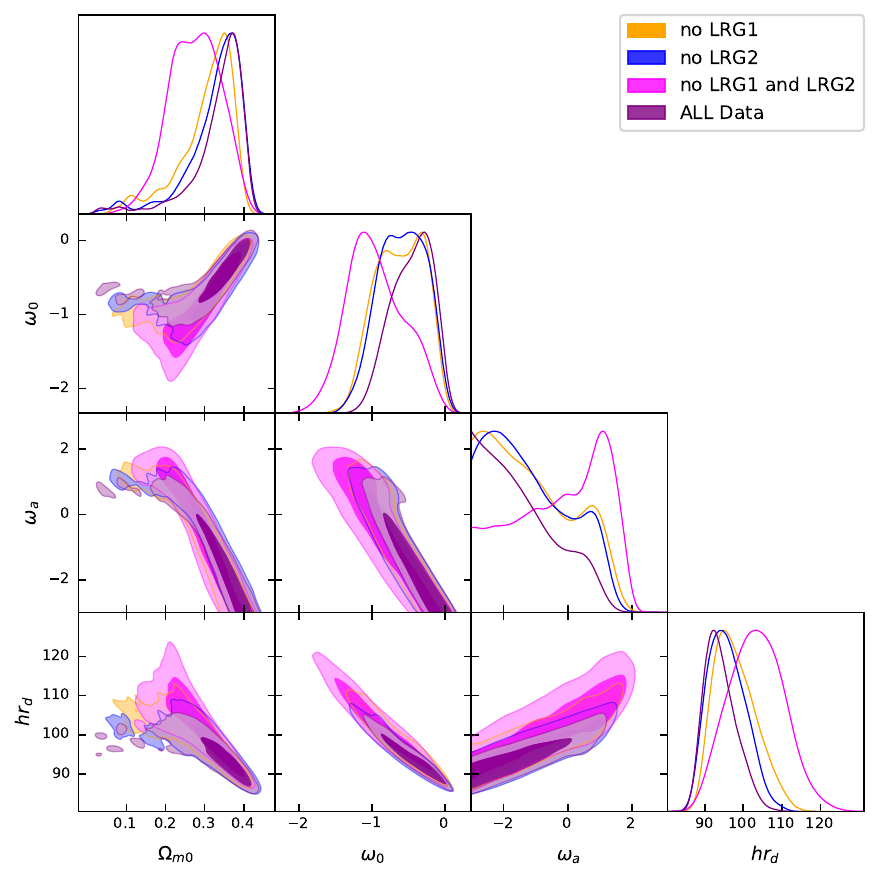}
\caption{The figure shows the confidence contours at the 68\% (1$\sigma$) and 95\% (2$\sigma$) levels for the $\omega_0 \omega_a$CDM model, using no LRG1, no LRG2, no LRG1 \& LRG2, and the full DESI DR2 sample.}\label{fig_5}
\end{figure}
\begin{figure}
\centering
\includegraphics[scale=0.5]{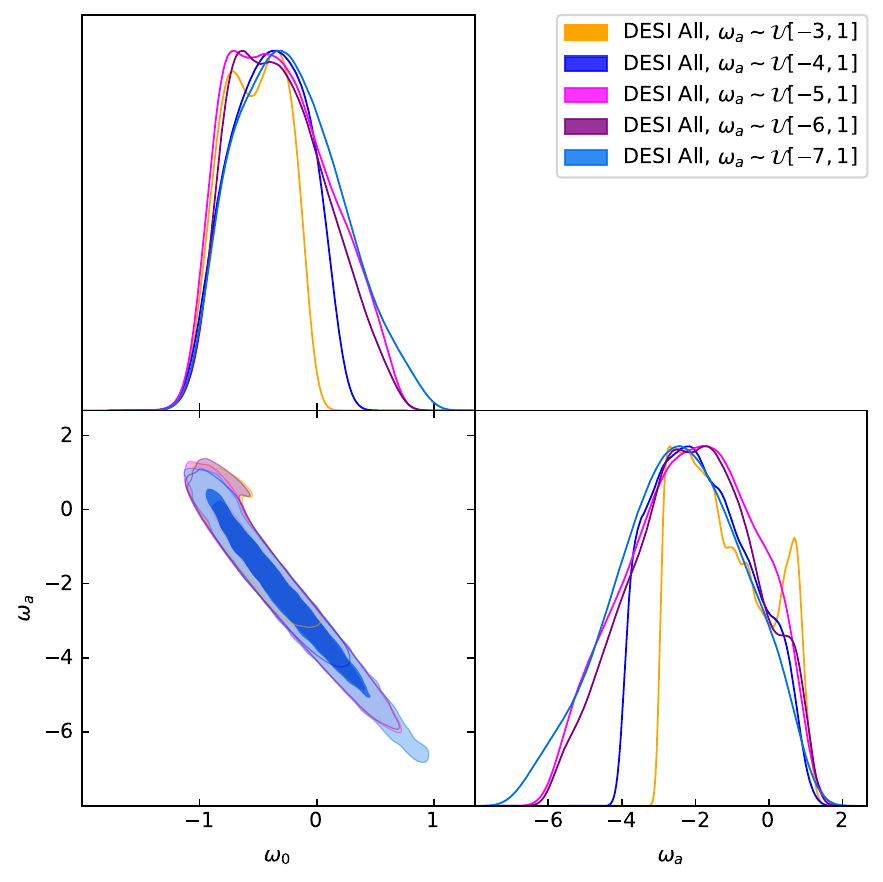}
\caption{The figure shows the Prior sensitivity of the $\omega_0\omega_a$CDM constraints from DESI DR2.}\label{fig_6}
\end{figure}
\begin{table}
\centering
\begin{tabular}{llcc}
\hline
\textbf{Parameter} & \textbf{Dataset} & \textbf{$\Lambda$CDM} & \textbf{$\omega_0 \omega_a$CDM} \\
\hline
& No LRG1 & $0.287_{\pm 0.007}^{\pm 0.013}$ & $0.33_{\pm 0.02}^{\pm 0.16}$ \\
$\Omega_{m0}$ & No LRG2 & $0.298_{\pm 0.006}^{\pm 0.013}$ & $0.36_{\pm 0.03}^{\pm 0.08}$ \\
& No LRG1 \& LRG2 & $0.291_{\pm 0.008}^{\pm 0.015}$ & $0.26_{\pm 0.06}^{\pm 0.14}$ \\
& DESI DR2 & $0.297 _{\pm 0.007}^{\pm 0.013}$ & $0.36_{\pm 0.02}^{\pm 0.03}$ \\
\hline
& No LRG1 & --- & $-0.52_{\pm 0.31}^{\pm 0.66}$ \\
$\omega_{0}$ & No LRG2 & --- & $-0.46_{\pm 0.29}^{\pm 0.61}$ \\
& No LRG1 \& LRG2 & --- & $-0.99_{\pm 0.37}^{\pm 0.63}$ \\
& DESI DR2 & --- & $-0.41_{\pm 0.20}^{\pm 0.48}$ \\
\hline
& No LRG1 & --- & $-1.66_{\pm 0.92}^{\pm 1.24}$ \\
$\omega_{a}$ & No LRG2 & --- & $-1.71_{\pm 0.97}^{\pm 1.22}$ \\
& No LRG1 \& LRG2 & --- & $-0.19_{\pm 1.69}^{\pm 2.60}$ \\
& DESI DR2 & --- & $-1.99_{\pm 0.69}^{\pm 0.93}$ \\
\hline
& No LRG1 & $102.59_{\pm 0.70}^{\pm 1.44}$ & $96.69_{\pm 4.12}^{\pm 6.67}$ \\
$h r_d$ & No LRG2 & $101.20_{\pm 0.54}^{\pm 1.30}$ & $93.93_{\pm 3.77}^{+5.}$ \\
& No LRG1 \& LRG2 & $102.04_{\pm 0.83}^{\pm 1.75}$ & $104.77_{\pm 7.29}^{\pm 11.36}$ \\
& DESI DR2 & $101.79_{\pm 0.61}^{\pm 1.17}$ & $93.29_{\pm 2.42}^{\pm 3.79}$ \\
\hline
& No LRG1 & 0 & 2.41 \\
$\ln$BF & No LRG2 & 0 & 2.06 \\
& No LRG1 \& LRG2 & 0 & 1.89 \\
& DESI DR2 & 0 & 0.10 \\
\hline
\end{tabular}
\caption{Constraints on the parameters of the $\Lambda$CDM and $\omega_0\omega_a$CDM models from DESI DR2 at 68\% (1$\sigma$) and 95\% (2$\sigma$) confidence levels, based on the full dataset as well as cases excluding the LRG1 and/or LRG2 redshift data points.}
\label{tab_4}
\end{table}

In this subsection, we constrain the parameters of the $\omega_0 \omega_a$CDM model using the DESI DR2 dataset alone. Fig.~\ref{fig_5} shows the constraints on the $\omega_0 \omega_a$CDM model using the DESI DR2 dataset, both with and without the LRG1 and LRG2 datasets, as well as excluding both simultaneously. The corresponding numerical results from the MCMC analysis are summarized in Table~\ref{tab_4}. When we exclude the LRG1 dataset, the $\omega_0\omega_a$CDM model yields $\omega_0 = -0.52 \pm 0.31$. A similar trend can be observed when \cite{colgain2024does} considers the DESI DR1 dataset. They obtain $\omega_0 = -0.560^{+0.266}_{-0.384}$ with the LRG1 dataset included, and $\omega_0 = -0.984^{+0.427}_{-0.422}$ when the LRG1 sample is excluded. Removing the LRG2 dataset while keeping LRG1 gives $\omega_0 = -0.46 \pm 0.29$. When both LRG1 and LRG2 are excluded, we obtain $\omega_0 = -0.99 \pm 0.37$, and for the full DESI DR2 compilation, we find $\omega_0 = -0.41 \pm 0.20$. It is important to note that whenever LRG1 and/or LRG2 are included, the inferred $\omega_0$ deviates significantly from the $\Lambda$CDM prediction ($\omega_0 = -1$). By contrast, when both LRG1 and LRG2 are removed, the estimate shifts back toward $\omega_0 = -1$, fully restoring concordance with the $\Lambda$CDM model.

These results provide three key insights. First, DESI DR2 shows deviations from $\Lambda$CDM concordance, suggesting the possibility of evolving dark energy with $\omega_0 > -1$. Second, this apparent evolution is largely driven by the LRG1 and LRG2 samples. Third, since $\omega_0^{\text{LRG2}} > \omega_0^{,\text{LRG1}}$, LRG2 appears to contribute more strongly than LRG1 to the dynamical dark energy signal. In \cite{colgain2025much}, Fig. 4 shows the faded blue curve corresponding to DESI DR2 predictions. It shows that in DESI DR1 the dynamical dark energy signal is driven by the combination of a higher $\Omega_m$ from LRG1, a lower $\Omega_m$ from LRG2, and an increase in $\Omega_m$ at higher redshifts. A similar pattern is indeed observed in the DR2 data, though shifted to higher redshift. The role of LRG2 is also discussed further in the appendix of \cite{goldstein2025monodromic}.

In Fig.~\ref{fig_6}, we also show that, when the full DESI DR2 data are used and the prior on $\omega_a$ is widened, in each case $\omega_a$ is pushed to large negative values exceeding the prior limits in order to accommodate $\omega_0>-1$ (see also \cite{cortes2024interpreting}). Recent studies likewise find that relaxing the prior on $\omega_a$ can drive $\omega_0$ to larger values, up to $\omega_0\sim 1$ \cite{wang2024self}, which is incompatible with late-time accelerated expansion ($\omega(z)<-1/3$). This behavior indicates sensitivity to prior volume along the $\omega_0$–$\omega_a$ degeneracy under limited per-tracer information; therefore, the apparent preference for $\omega_0>-1$ should not be interpreted as a data-driven detection. To assess whether the extra freedom is warranted, we report the (natural) log Bayes factor, $\ln \mathrm{BF}\equiv \ln\!\big(\mathcal{Z}_{\Lambda\mathrm{CDM}}/\mathcal{Z}_{\omega_0\omega_a\mathrm{CDM}}\big)$, computed with the \emph{same} likelihood and nuisance parameter priors for both models (the only extension being $\omega_a$). Following the Jeffreys scale \cite{jeffreys1998theory}, we interpret $|\ln \mathrm{BF}|<1$ as inconclusive, $1\le|\ln \mathrm{BF}|<2.5$ as weak evidence, $2.5\le|\ln \mathrm{BF}|<5$ as moderate evidence, and $|\ln \mathrm{BF}|\ge 5$ as strong evidence. For our DESI DR2 splits we find $\ln\mathrm{BF}={2.41,2.06,1.89,0.10}$ for {No LRG1, No LRG2, No LRG1\&LRG2, full DR2}, indicating weak evidence for $\Lambda$CDM in the first three cases and inconclusive support for the full sample. Hence the data do not require the additional $\omega_a$ degree of freedom; the apparent trend toward $\omega_0>-1$ with LRG1 included is best understood as prior volume sensitivity along the $\omega_0$–$\omega_a$ ridge rather than a data driven detection.

Our analysis uses the DESI BAO data only, so we can clearly see how each tracer affects the parameters. With DESI DR2 BAO alone, the $\omega_0\omega_a$CDM model is not well constrained. Stronger constraints require adding complementary data for example, combining DESI DR2 with Type Ia supernovae and the CMB (DESI DR2 + SNe Ia + CMB). Nevertheless, the predicted values of $\Omega_m$ in the LRG1 and LRG3+ELG1 tracers show strong disagreement with the Planck predictions, and when the LRG1 data point is included, DESI DR2 shows close agreement with the Planck–$\Lambda$CDM $\Omega_m$ predictions. One can conclude that the dynamic dark energy observed in the DESI DR2 dataset is primarily influenced by the LRG1 and LRG2 datasets. Furthermore, the DESI DR2 paper provides strong evidence for dynamic dark energy. When the DESI DR2 dataset is combined with other data sets, such as CMB, Pantheon$^+$, DESY5, and Union3 datasets, seems to be partly due to a mismatch between BAO and SNe Ia measurements at lower redshifts ($z_{\text{eff}} \sim 0.3$). Until the source of this discrepancy is better understood, claims of dynamic dark energy should be considered premature.

\section{Discussion and Conclusions}\label{sec_5}
The recent findings from the DESI DR2 data release suggest that the $\Lambda$CDM model is challenged by the combination of DESI DR2 BAO measurements and other observational data, motivating dynamical dark energy as a potential solution. Specifically, Fig.~11 of \cite{karim2025desi} shows that combining DESI DR2 with CMB data alone excludes $\Lambda$CDM at a significance of $3.1\sigma$. When additional supernova datasets are included, the exclusion significances become $2.8\sigma$ with Pantheon$^+$, $3.8\sigma$ with Union3, and $4.2\sigma$ with DES-SN5YR. Similarly, Fig.~\ref{fig_5} highlights a preference for $\omega_0 > -1$ within the $\omega_0\omega_a$CDM model when the LRG1 and LRG2 datasets are included, showing that these two tracers in DESI DR2 drive the preference for $\omega_0 > -1$.

The predicted value of $\Omega_m$ from DESI DR2 at $z_{\text{eff}} = 0.510$ (LRG1) shows a tension of about $2.42\sigma$ with the $\Lambda$CDM Planck prediction \cite{aghanim2020planck}. Similarly, the LRG3+ELG1 sample exhibits a $2.60\sigma$ tension relative to the Planck $\Omega_m$ value. These discrepancies place the DESI LRG1 and LRG3+ELG1 constraints in conflict not only with Planck, but also with several Type Ia supernova samples \cite{brout2022pantheon,rubin2025union,abbott2024dark}. Moreover, all tracers except Ly$\alpha$ exhibit some level of tension. It is important to note that the effective redshift is quite similar across these datasets, although SNe Ia samples tend to be biased towards lower redshifts.

Furthermore, Ref. \cite{karim2025desi} shows that SDSS and DESI constraints are consistent with each other. Consequently, SDSS-IV also exhibits a similar trend, predicting a larger $\Omega_m$ within the same range \cite{o2022revealing} (see Fig. 5). Although the predicted value of $\Omega_m$ at $z_{\text{eff}} = 0.51$ in the case of SDSS and DESI DR2 is very close, differences in $\Omega_m$ values between SDSS and DESI are noticeable at lower redshifts. While these differences may be attributed to systematics, they also suggest the possibility of statistical fluctuations in the data.

The success of the $\Lambda$CDM model is often highlighted by the consistent agreement across multiple observables, CMB, BAO, and SNe Ia, indicating that the Universe is composed of approximately 70\% dark energy and 30\% of matter. In cosmology, it is essential that physical models are supported by a variety of independent observations; otherwise, the results may simply reflect statistical fluctuations or observational systematics. This concern is particularly relevant in the case of the LRG1 and LRG3+ELG1 constraints at $z_{\text{eff}} = 0.510$ and $z_{\text{eff}} = 0.934$, as well as the three SNe Ia samples. It should be noted that these SNe Ia samples are not fully independent from each other, raising the possibility that systematic effects persist across the samples. For instance, Pantheon$^+$ and Union3 share approximately 1360 SNe Ia, and Pantheon$^+$ and DES-SN5YR share 196 SNe Ia.

The key takeaway from this paper is that while it is possible to compare the behavior of different models, such as the $\omega_0 \omega_a$CDM and $\Lambda$CDM models, we must ensure that the differences between these models are consistent across different types of measurements within the same redshift ranges. Without this consistency, claims of  new physics remain unconvincing, regardless of the attention they may attract \cite{tada2024quintessential,gu2024dynamical,wang2024dynamical,wang2024constraining,luongo2024model,yin2024cosmic}

As shown in Table~\ref{tab_2}, the DESI DR2 dataset exhibits a shift in the value of $\Omega_m$ of approximately 1.84$\sigma$ as we move from lower to higher redshift bins. Specifically, $\Omega_m$ decreases with increasing effective redshift bin before rising again. Consequently, the parameter $h r_d$ shows a negative correlation with $\Omega_m$, leading to similar behavior of increasing and then decreasing values. Type Ia supernova measurements also exhibit similar trends \cite{colgain2019hint,kazantzidis2021hints}. Indeed, this is not surprising, as large SNe Ia samples are typically compiled from several different surveys. As the number of surveys increases, observational systematics become a greater. However, recent studies have shown that the Pantheon$^+$ sample generally provides consistent $\Omega_m$ values \cite{brout2022pantheon}, provided that high redshift SNe Ia are excluded \cite{malekjani2024redshift}. In contrast, DESI is based on a single survey, though it incorporates multiple tracers.

The main idea here is that ongoing tensions with the $\Lambda$CDM model, particularly the discrepancies in $H_0$ and $S_8 = \sigma_8 \sqrt{\Omega_m / 0.3}$ \cite{di2021realm,perivolaropoulos2022challenges,abdalla2022cosmology}, suggest that the model may need to be revised if these tensions are found to be real. Such a revision could involve allowing cosmological parameters to evolve with effective redshift \cite{krishnan2021running,krishnan2023h}. While some may argue that the $S_8$ tension is merely a scale issue, as discussed in Refs. \cite{amon2022non,preston2023non}, this approach does not address the more statistically significant $H_0$ tension. Moreover, shifts in $S_8$ caused by changes in scale appear too small to fully resolve the $S_8$ tension \cite{terasawa2025exploring, DiValentino:2020vvd}. This reinforces the idea that the discrepancies in both $H_0$ and $S_8$ are not just scale effects, but may point to deeper issues within the $\Lambda$CDM model. Supporting evidence for this notion comes from various observations showing that $H_0$ decreases and $\Omega_m$ increases with increasing effective redshift, a trend reported in multiple studies \cite{o2022revealing,wong2020h0licow,wong2020h0licow,shajib2020strides,millon2020tdcosmo,kelly2023constraints,pascale2025sn,dainotti2021hubble,colgain2024putting,jia2023evidence,pasten2023testing,malekjani2024redshift,wagner2022solving,hu2022revealing,dainotti2023hubble,dainotti2022evolution,krishnan2020there,dainotti2022gamma,risaliti2019cosmological,lusso2020quasars,pascale2025sn}. If these findings hold, they could indicate that the $\Lambda$CDM model breaks down at the background level in the late Universe. Additionally, observations suggest that the $S_8$ tension, which seems to be a perturbative issue, is primarily located in the late Universe, especially at $z \lesssim 2$ \cite{esposito2022weighing,adil2024s,madhavacheril2024atacama,tutusaus2024measurement,CosmoVerse:2025txj,Abdalla:2022yfr}.. For more details on the evidence supporting the evolution of $\Lambda$CDM parameters with redshift, refer to \cite{akarsu2024lambda}.

DESI DR2 shows some improvement when compared to previous data, but the situation remains uncertain due to the persistence of the anomaly in the LRG1 ($z_{\text{eff}} = 0.51$) and LRG3+ELG1 ($z_{\text{eff}} = 0.943$) datasets. This anomaly could lead to discrepancies between BAO, CMB + SNe Ia measurements, and possibly even between CMB and SNe Ia data \cite{abbott2024dark} particularly regarding $\Omega_m$. On the other hand, if the data quality improves and the tension with Planck, Pantheon$^+$, and other measurements is reduced, it’s possible that the negative correlation between $\Omega_m$ and $h r_d$ at lower redshifts could disappear  Instead, we may observe the more typical trend of increasing $\Omega_m$ and decreasing $H_0$, which reflects a negative correlation between $\Omega_m$ and $H_0$. In a follow-up paper, we aim to explore various dark energy models and check whether the DESI LRG1 analogy holds and can be improved in all these models.

In our next paper, we intend to extend the above analysis and investigate these effects in greater depth, as it is important to recognize the statistical limitations of tracer-wise inference. Each DESI tracer provides only a small number of BAO observables (e.g., $D_H/r_d$, $D_M/r_d$, $D_V/r_d$, $D_M/D_H$), while models such as $\Lambda$CDM and $\omega_0\omega_a$CDM involve several free parameters. This imbalance makes individual tracer fits inherently underconstrained and prone to amplifying parameter degeneracies, particularly among $\Omega_m$, $H_0$, and $r_d$. Apparent redshift dependent trends or tensions may therefore arise from statistical fluctuations or prior volume effects rather than genuine signs of new physics. A more robust picture emerges only when combining tracers and cross-checking with complementary probes such as the CMB or Type Ia supernovae, which we plan to pursue in our upcoming work.
\section*{Acknowledgements}
SC acknowledges the Istituto Nazionale di Fisica Nucleare (INFN) Sez. di Napoli,  Iniziative Specifiche QGSKY and MoonLight-2  and the Istituto Nazionale di Alta Matematica (INdAM), gruppo GNFM, for the support.
This paper is based upon work from COST Action CA21136 -- Addressing observational tensions in cosmology with systematics and fundamental physics (CosmoVerse), supported by COST (European Cooperation in Science and Technology). Vipin K. Sharma gratefully acknowledges the facilities and institutional support provided by the Indian Institute of Astrophysics (IIA), India, during his tenure as a postdoctoral fellow.
\bibliographystyle{elsarticle-num}
\bibliography{mybib.bib}

\begin{thebibliography}{100}
\expandafter\ifx\csname url\endcsname\relax
  \def\url#1{\texttt{#1}}\fi
\expandafter\ifx\csname urlprefix\endcsname\relax\def\urlprefix{URL }\fi
\expandafter\ifx\csname href\endcsname\relax
  \def\href#1#2{#2} \def\path#1{#1}\fi

\bibitem{SupernovaSearchTeam:1998fmf}
A.~G. Riess, et~al., {Observational evidence from supernovae for an accelerating universe and a cosmological constant}, Astron. J. 116 (1998) 1009--1038.
\newblock \href {http://arxiv.org/abs/astro-ph/9805201} {\path{arXiv:astro-ph/9805201}}, \href {https://doi.org/10.1086/300499} {\path{doi:10.1086/300499}}.

\bibitem{SupernovaCosmologyProject:1998vns}
S.~Perlmutter, et~al., {Measurements of $\Omega$ and $\Lambda$ from 42 High Redshift Supernovae}, Astrophys. J. 517 (1999) 565--586.
\newblock \href {http://arxiv.org/abs/astro-ph/9812133} {\path{arXiv:astro-ph/9812133}}, \href {https://doi.org/10.1086/307221} {\path{doi:10.1086/307221}}.

\bibitem{Bamba:2012cp}
K.~Bamba, S.~Capozziello, S.~Nojiri, S.~D. Odintsov, {Dark energy cosmology: the equivalent description via different theoretical models and cosmography tests}, Astrophys. Space Sci. 342 (2012) 155--228.
\newblock \href {http://arxiv.org/abs/1205.3421} {\path{arXiv:1205.3421}}, \href {https://doi.org/10.1007/s10509-012-1181-8} {\path{doi:10.1007/s10509-012-1181-8}}.

\bibitem{Sousa-Neto:2025gpj}
A.~Sousa-Neto, C.~Bengaly, J.~E. Gonzalez, J.~Alcaniz, {Evidence for dynamical dark energy from DESI-DR2 and SN data? A symbolic regression analysis} (6 2025).
\newblock \href {http://arxiv.org/abs/2502.10506} {\path{arXiv:2502.10506}}.

\bibitem{Notari:2024zmi}
A.~Notari, M.~Redi, A.~Tesi, {BAO vs. SN evidence for evolving dark energy}, JCAP 04 (2025) 048.
\newblock \href {http://arxiv.org/abs/2411.11685} {\path{arXiv:2411.11685}}, \href {https://doi.org/10.1088/1475-7516/2025/04/048} {\path{doi:10.1088/1475-7516/2025/04/048}}.

\bibitem{Sharma:2025qmv}
V.~k. Sharma, H.~Chaudhary, S.~Kolekar, {Probing Generalized Emergent Dark Energy with DESI DR2} (7 2025).
\newblock \href {http://arxiv.org/abs/2507.00835} {\path{arXiv:2507.00835}}.

\bibitem{Weinberg:1988cp}
S.~Weinberg, {The Cosmological Constant Problem}, Rev. Mod. Phys. 61 (1989) 1--23.
\newblock \href {https://doi.org/10.1103/RevModPhys.61.1} {\path{doi:10.1103/RevModPhys.61.1}}.

\bibitem{YaBZeldovich_1968}
Y.~B. Zel'dovich, {The Cosmological Constant And The Theory Of Elementary Particles}, Soviet Physics Uspekhi 11~(3) (1968) 381.
\newblock \href {https://doi.org/10.1070/PU1968v011n03ABEH003927} {\path{doi:10.1070/PU1968v011n03ABEH003927}}.

\bibitem{DESI:2024mwx}
A.~G. Adame, et~al., {DESI 2024 VI: cosmological constraints from the measurements of baryon acoustic oscillations}, JCAP 02 (2025) 021.
\newblock \href {http://arxiv.org/abs/2404.03002} {\path{arXiv:2404.03002}}, \href {https://doi.org/10.1088/1475-7516/2025/02/021} {\path{doi:10.1088/1475-7516/2025/02/021}}.

\bibitem{DESI:2024aqx}
R.~Calderon, et~al., {DESI 2024: reconstructing dark energy using crossing statistics with DESI DR1 BAO data}, JCAP 10 (2024) 048.
\newblock \href {http://arxiv.org/abs/2405.04216} {\path{arXiv:2405.04216}}, \href {https://doi.org/10.1088/1475-7516/2024/10/048} {\path{doi:10.1088/1475-7516/2024/10/048}}.

\bibitem{karim2025desi}
M.~A. Karim, J.~Aguilar, S.~Ahlen, S.~Alam, L.~Allen, C.~Allende~Prieto, O.~Alves, A.~Anand, U.~Andrade, E.~Armengaud, et~al., Desi dr2 results ii: Measurements of baryon acoustic oscillations and cosmological constraints, arXiv e-prints (2025) arXiv--2503.

\bibitem{DESI:2024kob}
K.~Lodha, et~al., {DESI 2024: Constraints on physics-focused aspects of dark energy using DESI DR1 BAO data}, Phys. Rev. D 111~(2) (2025) 023532.
\newblock \href {http://arxiv.org/abs/2405.13588} {\path{arXiv:2405.13588}}, \href {https://doi.org/10.1103/PhysRevD.111.023532} {\path{doi:10.1103/PhysRevD.111.023532}}.

\bibitem{Sahni:2006pa}
V.~Sahni, A.~Starobinsky, {Reconstructing Dark Energy}, Int. J. Mod. Phys. D 15 (2006) 2105--2132.
\newblock \href {http://arxiv.org/abs/astro-ph/0610026} {\path{arXiv:astro-ph/0610026}}, \href {https://doi.org/10.1142/S0218271806009704} {\path{doi:10.1142/S0218271806009704}}.

\bibitem{Chevallier:2000qy}
M.~Chevallier, D.~Polarski, {Accelerating universes with scaling dark matter}, Int. J. Mod. Phys. D 10 (2001) 213--224.
\newblock \href {http://arxiv.org/abs/gr-qc/0009008} {\path{arXiv:gr-qc/0009008}}, \href {https://doi.org/10.1142/S0218271801000822} {\path{doi:10.1142/S0218271801000822}}.

\bibitem{Linder:2002et}
E.~V. Linder, {Exploring the expansion history of the universe}, Phys. Rev. Lett. 90 (2003) 091301.
\newblock \href {http://arxiv.org/abs/astro-ph/0208512} {\path{arXiv:astro-ph/0208512}}, \href {https://doi.org/10.1103/PhysRevLett.90.091301} {\path{doi:10.1103/PhysRevLett.90.091301}}.

\bibitem{Park:2024pew}
C.-G. Park, J.~de~Cruz~P{\'e}rez, B.~Ratra, {Is the $w_0w_a$CDM cosmological parameterization evidence for dark energy dynamics partially caused by the excess smoothing of Planck CMB anisotropy data?} (10 2024).
\newblock \href {http://arxiv.org/abs/2410.13627} {\path{arXiv:2410.13627}}.

\bibitem{wolf2025matching}
W.~J. Wolf, P.~G. Ferreira, C.~Garc{\'\i}a-Garc{\'\i}a, Matching current observational constraints with nonminimally coupled dark energy, Physical Review D 111~(4) (2025) L041303.

\bibitem{wolf2025robustness}
W.~J. Wolf, C.~Garc{\'\i}a-Garc{\'\i}a, P.~G. Ferreira, Robustness of dark energy phenomenology across different parameterizations, Journal of Cosmology and Astroparticle Physics 2025~(05) (2025) 034.

\bibitem{Vilardi:2024cwq}
S.~Vilardi, S.~Capozziello, M.~Brescia, {Discriminating between cosmological models using data-driven methods}, Astron. Astrophys. 695 (2025) A166.
\newblock \href {http://arxiv.org/abs/2408.01563} {\path{arXiv:2408.01563}}, \href {https://doi.org/10.1051/0004-6361/202451779} {\path{doi:10.1051/0004-6361/202451779}}.

\bibitem{Chudaykin:2020ghx}
A.~Chudaykin, K.~Dolgikh, M.~M. Ivanov, {Constraints on the curvature of the Universe and dynamical dark energy from the Full-shape and BAO data}, Phys. Rev. D 103~(2) (2021) 023507.
\newblock \href {http://arxiv.org/abs/2009.10106} {\path{arXiv:2009.10106}}, \href {https://doi.org/10.1103/PhysRevD.103.023507} {\path{doi:10.1103/PhysRevD.103.023507}}.

\bibitem{Escamilla-Rivera:2019aol}
C.~Escamilla-Rivera, S.~Capozziello, {Unveiling cosmography from the dark energy equation of state}, Int. J. Mod. Phys. D 28~(12) (2019) 1950154.
\newblock \href {http://arxiv.org/abs/1905.04602} {\path{arXiv:1905.04602}}, \href {https://doi.org/10.1142/S0218271819501542} {\path{doi:10.1142/S0218271819501542}}.

\bibitem{handley2015polychord1}
W.~Handley, M.~Hobson, A.~Lasenby, Polychord: next-generation nested sampling, Monthly Notices of the Royal Astronomical Society 453~(4) (2015) 4384--4398.

\bibitem{handley2015polychord2}
W.~Handley, M.~Hobson, A.~Lasenby, Polychord: nested sampling for cosmology, Monthly Notices of the Royal Astronomical Society: Letters 450~(1) (2015) L61--L65.

\bibitem{lewis2019getdist}
A.~Lewis, Getdist: a python package for analysing monte carlo samples, arXiv preprint arXiv:1910.13970 (2019).

\bibitem{pogosian2020recombination}
L.~Pogosian, G.-B. Zhao, K.~Jedamzik, Recombination-independent determination of the sound horizon and the hubble constant from bao, The Astrophysical Journal Letters 904~(2) (2020) L17.

\bibitem{jedamzik2021reducing}
K.~Jedamzik, L.~Pogosian, G.-B. Zhao, Why reducing the cosmic sound horizon alone can not fully resolve the hubble tension, Communications Physics 4~(1) (2021) 123.

\bibitem{pogosian2024consistency}
L.~Pogosian, G.-B. Zhao, K.~Jedamzik, A consistency test of the cosmological model at the epoch of recombination using desi bao and planck measurements, arXiv preprint arXiv:2405.20306 (2024).

\bibitem{lin2021early}
W.~Lin, X.~Chen, K.~J. Mack, Early-universe-physics insensitive and uncalibrated cosmic standards: Constraints on $\omega_{m}$ and implications for the hubble tension, arXiv preprint arXiv:2102.05701 (2021).

\bibitem{vagnozzi2023seven}
S.~Vagnozzi, Seven hints that early-time new physics alone is not sufficient to solve the hubble tension, Universe 9~(9) (2023) 393.

\bibitem{aghanim2020planck}
N.~Aghanim, Y.~Akrami, M.~Ashdown, J.~Aumont, C.~Baccigalupi, M.~Ballardini, A.~J. Banday, R.~Barreiro, N.~Bartolo, S.~Basak, et~al., Planck 2018 results-vi. cosmological parameters, Astronomy \& Astrophysics 641 (2020) A6.

\bibitem{colgain2024does}
E.~{\'O}. Colg{\'a}in, M.~G. Dainotti, S.~Capozziello, S.~Pourojaghi, M.~Sheikh-Jabbari, D.~Stojkovic, Does desi2024 confirm $\lambda$cdm?, Journal of High Energy Astrophysics 49 (2026) 100428.

\bibitem{brout2022pantheon}
D.~Brout, D.~Scolnic, B.~Popovic, A.~G. Riess, A.~Carr, J.~Zuntz, R.~Kessler, T.~M. Davis, S.~Hinton, D.~Jones, et~al., The pantheon+ analysis: cosmological constraints, The Astrophysical Journal 938~(2) (2022) 110.

\bibitem{rubin2025union}
D.~Rubin, G.~Aldering, M.~Betoule, A.~Fruchter, X.~Huang, A.~G. Kim, C.~Lidman, E.~Linder, S.~Perlmutter, P.~Ruiz-Lapuente, et~al., Union through unity: Cosmology with 2000 sne using a unified bayesian framework, The Astrophysical Journal 986~(2) (2025) 231.

\bibitem{abbott2024dark}
T.~Abbott, M.~Acevedo, M.~Aguena, A.~Alarcon, S.~Allam, O.~Alves, A.~Amon, F.~Andrade-Oliveira, J.~Annis, P.~Armstrong, et~al., The dark energy survey: Cosmology results with\~{} 1500 new high-redshift type ia supernovae using the full 5-year dataset, arXiv preprint arXiv:2401.02929 (2024).

\bibitem{o2022revealing}
E.~{\'O}~Colg{\'a}in, M.~Sheikh-Jabbari, R.~Solomon, G.~Bargiacchi, S.~Capozziello, M.~G. Dainotti, D.~Stojkovic, Revealing intrinsic flat $\lambda$ cdm biases with standardizable candles, Physical Review D 106~(4) (2022) L041301.

\bibitem{eisenstein2005detection}
D.~J. Eisenstein, et~al., Detection of the baryon acoustic peak in the large-scale correlation function of sdss luminous red galaxies, The Astrophysical Journal 633~(2) (2005) 560.
\newblock \href {https://doi.org/10.1086/466512} {\path{doi:10.1086/466512}}.

\bibitem{percival2010baryon}
W.~J. Percival, et~al., Baryon acoustic oscillations in the sloan digital sky survey data release 7 galaxy sample, Monthly Notices of the Royal Astronomical Society 401~(4) (2010) 2148--2168.
\newblock \href {https://doi.org/10.1111/j.1365-2966.2009.15812.x} {\path{doi:10.1111/j.1365-2966.2009.15812.x}}.

\bibitem{aubourg2015cosmological}
{\'E}.~Aubourg, et~al., Cosmological implications of baryon acoustic oscillation (bao) measurements, Physical Review D 92~(12) (2015) 123516.
\newblock \href {https://doi.org/10.1103/PhysRevD.92.123516} {\path{doi:10.1103/PhysRevD.92.123516}}.

\bibitem{alam2017clustering}
S.~Alam, M.~Ata, S.~Bailey, F.~Beutler, D.~Bizyaev, J.~A. Blazek, A.~S. Bolton, J.~R. Brownstein, A.~Burden, C.-H. Chuang, et~al., The clustering of galaxies in the completed sdss-iii baryon oscillation spectroscopic survey: cosmological analysis of the dr12 galaxy sample, Monthly Notices of the Royal Astronomical Society 470~(3) (2017) 2617--2652.

\bibitem{alam2021completed}
S.~Alam, M.~Aubert, S.~Avila, C.~Balland, J.~E. Bautista, M.~A. Bershady, D.~Bizyaev, M.~R. Blanton, A.~S. Bolton, J.~Bovy, et~al., Completed sdss-iv extended baryon oscillation spectroscopic survey: Cosmological implications from two decades of spectroscopic surveys at the apache point observatory, Physical Review D 103~(8) (2021) 083533.

\bibitem{colgain2024putting}
E.~{\'O}. Colg{\'a}in, M.~Sheikh-Jabbari, R.~Solomon, M.~G. Dainotti, D.~Stojkovic, Putting flat $\lambda$cdm in the (redshift) bin, Physics of the Dark Universe 44 (2024) 101464.

\bibitem{dainotti2022evolution}
M.~G. Dainotti, B.~De~Simone, T.~Schiavone, G.~Montani, E.~Rinaldi, G.~Lambiase, M.~Bogdan, S.~Ugale, On the evolution of the hubble constant with the sne ia pantheon sample and baryon acoustic oscillations: a feasibility study for grb-cosmology in 2030, Galaxies 10~(1) (2022) 24.

\bibitem{dainotti2021hubble}
M.~G. Dainotti, B.~De~Simone, T.~Schiavone, G.~Montani, E.~Rinaldi, G.~Lambiase, On the hubble constant tension in the sne ia pantheon sample, The Astrophysical Journal 912~(2) (2021) 150.

\bibitem{jia2023evidence}
X.~Jia, J.~Hu, F.~Wang, Evidence of a decreasing trend for the hubble constant, Astronomy \& Astrophysics 674 (2023) A45.

\bibitem{pasten2023testing}
E.~Past{\'e}n, V.~H. C{\'a}rdenas, Testing $\lambda$cdm cosmology in a binned universe: Anomalies in the deceleration parameter, Physics of the Dark Universe 40 (2023) 101224.

\bibitem{malekjani2024redshift}
M.~Malekjani, R.~Mc~Conville, E.~{\'O}~Colg{\'a}in, S.~Pourojaghi, M.~Sheikh-Jabbari, On redshift evolution and negative dark energy density in pantheon+ supernovae, The European Physical Journal C 84~(3) (2024) 317.

\bibitem{wagner2022solving}
J.~Wagner, Solving the hubble tension$\backslash$a la ellis \& stoeger 1987, arXiv preprint arXiv:2203.11219 (2022).

\bibitem{hu2022revealing}
J.-P. Hu, F.-Y. Wang, Revealing the late-time transition of h 0: relieve the hubble crisis, Monthly Notices of the Royal Astronomical Society 517~(1) (2022) 576--581.

\bibitem{dainotti2023hubble}
M.~Dainotti, B.~De~Simone, G.~Montani, T.~Schiavone, G.~Lambiase, The hubble constant tension: current status and future perspectives through new cosmological probes, arXiv preprint arXiv:2301.10572 (2023).

\bibitem{krishnan2020there}
C.~Krishnan, E.~{\'O}. Colg{\'a}in, Ruchika, A.~A. Sen, M.~Sheikh-Jabbari, T.~Yang, Is there an early universe solution to hubble tension?, Physical Review D 102~(10) (2020) 103525.

\bibitem{dainotti2022gamma}
M.~G. Dainotti, G.~Sarracino, S.~Capozziello, Gamma-ray bursts, supernovae ia, and baryon acoustic oscillations: A binned cosmological analysis, Publications of the Astronomical Society of Japan 74~(5) (2022) 1095--1113.

\bibitem{Bargiacchi:2024srw}
G.~Bargiacchi, M.~G. Dainotti, S.~Capozziello, {High-redshift cosmology by Gamma-Ray Bursts: An overview}, New Astron. Rev. 100 (2025) 101712.
\newblock \href {http://arxiv.org/abs/2408.10707} {\path{arXiv:2408.10707}}, \href {https://doi.org/10.1016/j.newar.2024.101712} {\path{doi:10.1016/j.newar.2024.101712}}.

\bibitem{risaliti2019cosmological}
G.~Risaliti, E.~Lusso, Cosmological constraints from the hubble diagram of quasars at high redshifts, Nature Astronomy 3~(3) (2019) 272--277.

\bibitem{lusso2020quasars}
E.~Lusso, G.~Risaliti, E.~Nardini, G.~Bargiacchi, M.~Benetti, S.~Bisogni, S.~Capozziello, F.~Civano, L.~Eggleston, M.~Elvis, et~al., Quasars as standard candles-iii. validation of a new sample for cosmological studies, Astronomy \& Astrophysics 642 (2020) A150.

\bibitem{dainotti2023quasars}
M.~G. Dainotti, G.~Bargiacchi, A.~{\L}. Lenart, S.~Nagataki, S.~Capozziello, Quasars: Standard candles up to z= 7.5 with the precision of supernovae ia, The Astrophysical Journal 950~(1) (2023) 45.

\bibitem{dainotti2022quasar}
M.~G. Dainotti, G.~Bargiacchi, A.~{\L}. Lenart, S.~Capozziello, E.~{\'O}. Colg{\'a}in, R.~Solomon, D.~Stojkovic, M.~Sheikh-Jabbari, Quasar standardization: overcoming selection biases and redshift evolution, The Astrophysical Journal 931~(2) (2022) 106.

\bibitem{bargiacchi2023gamma}
G.~Bargiacchi, M.~Dainotti, S.~Nagataki, S.~Capozziello, Gamma-ray bursts, quasars, baryonic acoustic oscillations, and supernovae ia: new statistical insights and cosmological constraints, Monthly Notices of the Royal Astronomical Society 521~(3) (2023) 3909--3924.

\bibitem{pourojaghi2022can}
S.~Pourojaghi, N.~Zabihi, M.~Malekjani, Can high-redshift hubble diagrams rule out the standard model of cosmology in the context of cosmography?, Physical Review D 106~(12) (2022) 123523.

\bibitem{wong2020h0licow}
K.~C. Wong, S.~H. Suyu, G.~C. Chen, C.~E. Rusu, M.~Millon, D.~Sluse, V.~Bonvin, C.~D. Fassnacht, S.~Taubenberger, M.~W. Auger, et~al., H0licow--xiii. a 2.4 per cent measurement of h 0 from lensed quasars: 5.3 $\sigma$ tension between early-and late-universe probes, Monthly Notices of the Royal Astronomical Society 498~(1) (2020) 1420--1439.

\bibitem{shajib2020strides}
A.~J. Shajib, S.~Birrer, T.~Treu, A.~Agnello, E.~Buckley-Geer, J.~Chan, L.~Christensen, C.~Lemon, H.~Lin, M.~Millon, et~al., Strides: a 3.9 per cent measurement of the hubble constant from the strong lens system des j0408- 5354, Monthly Notices of the Royal Astronomical Society 494~(4) (2020) 6072--6102.

\bibitem{millon2020tdcosmo}
M.~Millon, A.~Galan, F.~Courbin, T.~Treu, S.~Suyu, X.~Ding, S.~Birrer, G.-F. Chen, A.~Shajib, D.~Sluse, et~al., Tdcosmo-i. an exploration of systematic uncertainties in the inference of h0 from time-delay cosmography, Astronomy \& Astrophysics 639 (2020) A101.

\bibitem{kelly2023constraints}
P.~L. Kelly, S.~Rodney, T.~Treu, M.~Oguri, W.~Chen, A.~Zitrin, S.~Birrer, V.~Bonvin, L.~Dessart, J.~M. Diego, et~al., Constraints on the hubble constant from supernova refsdal’s reappearance, Science 380~(6649) (2023) eabh1322.

\bibitem{pascale2025sn}
M.~Pascale, B.~L. Frye, J.~D. Pierel, W.~Chen, P.~L. Kelly, S.~H. Cohen, R.~A. Windhorst, A.~G. Riess, P.~S. Kamieneski, J.~M. Diego, et~al., Sn h0pe: the first measurement of h0 from a multiply imaged type ia supernova, discovered by jwst, The Astrophysical Journal 979~(1) (2025) 13.

\bibitem{li2024determining}
X.~Li, K.~Liao, Determining cosmological-model-independent h 0 with gravitationally lensed supernova refsdal, The Astrophysical Journal 966~(1) (2024) 121.

\bibitem{khadka2020using}
N.~Khadka, B.~Ratra, Using quasar x-ray and uv flux measurements to constrain cosmological model parameters, Monthly Notices of the Royal Astronomical Society 497~(1) (2020) 263--278.

\bibitem{khadka2021determining}
N.~Khadka, B.~Ratra, Determining the range of validity of quasar x-ray and uv flux measurements for constraining cosmological model parameters, Monthly Notices of the Royal Astronomical Society 502~(4) (2021) 6140--6156.

\bibitem{singal2022x}
J.~Singal, S.~Mutchnick, V.~Petrosian, The x-ray luminosity function evolution of quasars and the correlation between the x-ray and ultraviolet luminosities, The Astrophysical Journal 932~(2) (2022) 111.

\bibitem{petrosian2022can}
V.~Petrosian, J.~Singal, S.~Mutchnick, Can the distance--redshift relation be determined from correlations between luminosities?, The Astrophysical Journal Letters 935~(1) (2022) L19.

\bibitem{zajavcek2024effect}
M.~Zaja{\v{c}}ek, B.~Czerny, N.~Khadka, M.~L. Mart{\'\i}nez-Aldama, R.~Prince, S.~Panda, B.~Ratra, Effect of extinction on quasar luminosity distances determined from uv and x-ray flux measurements, The Astrophysical Journal 961~(2) (2024) 229.

\bibitem{dainotti2024scavenger}
M.~G. Dainotti, G.~Bargiacchi, A.~{\L}. Lenart, S.~Capozziello, The scavenger hunt for quasar samples to be used as cosmological tools, Galaxies 12~(1) (2024) 4.

\bibitem{dainotti2024new}
M.~G. Dainotti, A.~Lenart, M.~G. Yengejeh, S.~Chakraborty, N.~Fraija, E.~Di~Valentino, G.~Montani, A new binning method to choose a standard set of quasars, Physics of the Dark Universe 44 (2024) 101428.

\bibitem{Bargiacchi:2021hdp}
G.~Bargiacchi, M.~Benetti, S.~Capozziello, E.~Lusso, G.~Risaliti, M.~Signorini, {Quasar cosmology: dark energy evolution and spatial curvature}, Mon. Not. Roy. Astron. Soc. 515~(2) (2022) 1795--1806.
\newblock \href {http://arxiv.org/abs/2111.02420} {\path{arXiv:2111.02420}}, \href {https://doi.org/10.1093/mnras/stac1941} {\path{doi:10.1093/mnras/stac1941}}.

\bibitem{Benetti:2025ljc}
M.~Benetti, G.~Bargiacchi, G.~Risaliti, S.~Capozziello, E.~Lusso, M.~Signorini, {Quasar cosmology II: Joint analyses with cosmic microwave background}, Phys. Dark Univ. 49 (2025) 101983.
\newblock \href {http://arxiv.org/abs/2506.21477} {\path{arXiv:2506.21477}}, \href {https://doi.org/10.1016/j.dark.2025.101983} {\path{doi:10.1016/j.dark.2025.101983}}.

\bibitem{lenart2023bias}
A.~{\L}. Lenart, G.~Bargiacchi, M.~G. Dainotti, S.~Nagataki, S.~Capozziello, A bias-free cosmological analysis with quasars alleviating h 0 tension, The Astrophysical Journal Supplement Series 264~(2) (2023) 46.

\bibitem{camarena2025designing}
D.~Camarena, K.~Greene, J.~Houghteling, F.-Y. Cyr-Racine, Designing concordant distances in the age of precision cosmology: the impact of density fluctuations, arXiv preprint arXiv:2507.17969 (2025).

\bibitem{demianski2017cosmology}
M.~Demianski, E.~Piedipalumbo, D.~Sawant, L.~Amati, Cosmology with gamma-ray bursts-i. the hubble diagram through the calibrated ep, i--eiso correlation, Astronomy \& Astrophysics 598 (2017) A112.

\bibitem{dainotti2023gamma}
M.~Dainotti, A.~{\L}. Lenart, A.~Chraya, G.~Sarracino, S.~Nagataki, N.~Fraija, S.~Capozziello, M.~Bogdan, The gamma-ray bursts fundamental plane correlation as a cosmological tool, Monthly Notices of the Royal Astronomical Society 518~(2) (2023) 2201--2240.

\bibitem{khadka2021gamma}
N.~Khadka, O.~Luongo, M.~Muccino, B.~Ratra, Do gamma-ray burst measurements provide a useful test of cosmological models?, Journal of Cosmology and Astroparticle Physics 2021~(09) (2021) 042.

\bibitem{alfano2024cosmological}
A.~C. Alfano, S.~Capozziello, O.~Luongo, M.~Muccino, Cosmological transition epoch from gamma-ray burst correlations, Journal of High Energy Astrophysics 42 (2024) 178--196.

\bibitem{dainotti2008time}
M.~G. Dainotti, V.~F. Cardone, S.~Capozziello, A time--luminosity correlation for $\gamma$-ray bursts in the x-rays, Monthly Notices of the Royal Astronomical Society: Letters 391~(1) (2008) L79--L83.

\bibitem{srinivasaragavan2020investigation}
G.~P. Srinivasaragavan, M.~G. Dainotti, N.~Fraija, X.~Hernandez, S.~Nagataki, A.~Lenart, L.~Bowden, R.~Wagner, On the investigation of the closure relations for gamma-ray bursts observed by swift in the post-plateau phase and the grb fundamental plane, The Astrophysical Journal 903~(1) (2020) 18.

\bibitem{akarsu2024lambda}
{\"O}.~Akarsu, E.~O. Colgain, A.~A. Sen, M.~Sheikh-Jabbari, $\lambda$ cdm tensions: Localising missing physics through consistency checks, arXiv preprint arXiv:2402.04767 (2024).

\bibitem{colgain2025much}
E.~{\'O}. Colg{\'a}in, S.~Pourojaghi, M.~Sheikh-Jabbari, L.~Yin, How much has desi dark energy evolved since dr1?, arXiv preprint arXiv:2504.04417 (2025).

\bibitem{goldstein2025monodromic}
S.~Goldstein, M.~Celoria, F.~Schmidt, Monodromic dark energy and desi, arXiv preprint arXiv:2507.16970 (2025).

\bibitem{cortes2024interpreting}
M.~Cort{\^e}s, A.~R. Liddle, Interpreting desi's evidence for evolving dark energy, Journal of Cosmology and Astroparticle Physics 2024~(12) (2024) 007.

\bibitem{wang2024self}
D.~Wang, The self-consistency of desi analysis and comment on" does desi 2024 confirm $\lambda$cdm ?", arXiv preprint arXiv:2404.13833 (2024).

\bibitem{jeffreys1998theory}
H.~Jeffreys, The theory of probability, OuP Oxford, 1998.

\bibitem{tada2024quintessential}
Y.~Tada, T.~Terada, Quintessential interpretation of the evolving dark energy in light of desi observations, Physical Review D 109~(12) (2024) L121305.

\bibitem{gu2024dynamical}
G.~Gu, X.~Wang, X.~Mu, S.~Yuan, G.-B. Zhao, Dynamical dark energy in light of cosmic distance measurements. i. a demonstration using simulated datasets, Research in Astronomy and Astrophysics 24~(6) (2024) 065001.

\bibitem{wang2024dynamical}
X.~Wang, G.~Gu, X.~Mu, S.~Yuan, G.-B. Zhao, Dynamical dark energy in light of cosmic distance measurements. ii. a study using current observations, Research in Astronomy and Astrophysics 24~(6) (2024) 065002.

\bibitem{wang2024constraining}
D.~Wang, Constraining cosmological physics with desi bao observations, arXiv preprint arXiv:2404.06796 (2024).

\bibitem{luongo2024model}
O.~Luongo, M.~Muccino, Model-independent cosmographic constraints from desi 2024, Astronomy \& Astrophysics 690 (2024) A40.

\bibitem{yin2024cosmic}
W.~Yin, Cosmic clues: Desi, dark energy, and the cosmological constant problem, Journal of High Energy Physics 2024~(5) (2024) 1--9.

\bibitem{colgain2019hint}
E.~{\'O}. Colg{\'a}in, A hint of matter underdensity at low z?, Journal of Cosmology and Astroparticle Physics 2019~(09) (2019) 006.

\bibitem{kazantzidis2021hints}
L.~Kazantzidis, H.~Koo, S.~Nesseris, L.~Perivolaropoulos, A.~Shafieloo, Hints for possible low redshift oscillation around the best-fitting $\lambda$cdm model in the expansion history of the universe, Monthly Notices of the Royal Astronomical Society 501~(3) (2021) 3421--3426.

\bibitem{di2021realm}
E.~Di~Valentino, O.~Mena, S.~Pan, L.~Visinelli, W.~Yang, A.~Melchiorri, D.~F. Mota, A.~G. Riess, J.~Silk, In the realm of the hubble tension—a review of solutions, Classical and Quantum Gravity 38~(15) (2021) 153001.

\bibitem{perivolaropoulos2022challenges}
L.~Perivolaropoulos, F.~Skara, Challenges for $\lambda$cdm: An update, New Astronomy Reviews 95 (2022) 101659.

\bibitem{abdalla2022cosmology}
E.~Abdalla, G.~F. Abell{\'a}n, A.~Aboubrahim, A.~Agnello, {\"O}.~Akarsu, Y.~Akrami, G.~Alestas, D.~Aloni, L.~Amendola, L.~A. Anchordoqui, et~al., Cosmology intertwined: A review of the particle physics, astrophysics, and cosmology associated with the cosmological tensions and anomalies, Journal of High Energy Astrophysics 34 (2022) 49--211.

\bibitem{krishnan2021running}
C.~Krishnan, E.~{\'O}~Colg{\'a}in, M.~Sheikh-Jabbari, T.~Yang, Running hubble tension and a h0 diagnostic, Physical Review D 103~(10) (2021) 103509.

\bibitem{krishnan2023h}
C.~Krishnan, R.~Mondol, H 0 as a universal flrw diagnostic, in: Cosmology Workshop on A Multipolar Universe, Springer, 2023, pp. 35--52.

\bibitem{amon2022non}
A.~Amon, G.~Efstathiou, A non-linear solution to the s 8 tension?, Monthly Notices of the Royal Astronomical Society 516~(4) (2022) 5355--5366.

\bibitem{preston2023non}
C.~Preston, A.~Amon, G.~Efstathiou, A non-linear solution to the s 8 tension--ii. analysis of des year 3 cosmic shear, Monthly Notices of the Royal Astronomical Society 525~(4) (2023) 5554--5564.

\bibitem{terasawa2025exploring}
R.~Terasawa, X.~Li, M.~Takada, T.~Nishimichi, S.~Tanaka, S.~Sugiyama, T.~Kurita, T.~Zhang, M.~Shirasaki, R.~Takahashi, et~al., Exploring the baryonic effect signature in the hyper suprime-cam year 3 cosmic shear two-point correlations on small scales: The s 8 tension remains present, Physical Review D 111~(6) (2025) 063509.

\bibitem{DiValentino:2020vvd}
E.~Di~Valentino, et~al., {Cosmology Intertwined III: $f \sigma_8$ and $S_8$}, Astropart. Phys. 131 (2021) 102604.
\newblock \href {http://arxiv.org/abs/2008.11285} {\path{arXiv:2008.11285}}, \href {https://doi.org/10.1016/j.astropartphys.2021.102604} {\path{doi:10.1016/j.astropartphys.2021.102604}}.

\bibitem{esposito2022weighing}
M.~Esposito, V.~Ir{\v{s}}i{\v{c}}, M.~Costanzi, S.~Borgani, A.~Saro, M.~Viel, Weighing cosmic structures with clusters of galaxies and the intergalactic medium, Monthly Notices of the Royal Astronomical Society 515~(1) (2022) 857--870.

\bibitem{adil2024s}
S.~A. Adil, {\"O}.~Akarsu, M.~Malekjani, E.~{\'O}~Colg{\'a}in, S.~Pourojaghi, A.~A. Sen, M.~Sheikh-Jabbari, S 8 increases with effective redshift in $\lambda$cdm cosmology, Monthly Notices of the Royal Astronomical Society: Letters 528~(1) (2024) L20--L26.

\bibitem{madhavacheril2024atacama}
M.~S. Madhavacheril, F.~J. Qu, B.~D. Sherwin, N.~MacCrann, Y.~Li, I.~Abril-Cabezas, P.~A. Ade, S.~Aiola, T.~Alford, M.~Amiri, et~al., The atacama cosmology telescope: Dr6 gravitational lensing map and cosmological parameters, The Astrophysical Journal 962~(2) (2024) 113.

\bibitem{tutusaus2024measurement}
I.~Tutusaus, C.~Bonvin, N.~Grimm, Measurement of the weyl potential evolution from the first three years of dark energy survey data, Nature Communications 15~(1) (2024) 9295.

\bibitem{CosmoVerse:2025txj}
E.~Di~Valentino, et~al., {The CosmoVerse White Paper: Addressing observational tensions in cosmology with systematics and fundamental physics} (4 2025).
\newblock \href {http://arxiv.org/abs/2504.01669} {\path{arXiv:2504.01669}}, \href {https://doi.org/10.1016/j.dark.2025.101965} {\path{doi:10.1016/j.dark.2025.101965}}.

\bibitem{Abdalla:2022yfr}
E.~Abdalla, et~al., {Cosmology intertwined: A review of the particle physics, astrophysics, and cosmology associated with the cosmological tensions and anomalies}, JHEAp 34 (2022) 49--211.
\newblock \href {http://arxiv.org/abs/2203.06142} {\path{arXiv:2203.06142}}, \href {https://doi.org/10.1016/j.jheap.2022.04.002} {\path{doi:10.1016/j.jheap.2022.04.002}}.

\end{thebibliography}

\end{document}